\documentclass[preprint,eqsecnum,floats,aps]{revtex4}
\usepackage{graphicx}
\usepackage{bm}
\begin{document}
\include{def}
\def\v#1{\mbox{\boldmath$#1$}}
\newcommand{\smod}{******START MODIFICATION****** : }
\def\ket#1{|#1 \rangle}
\def\bra#1{\langle #1|}

\draft
\vspace{2cm}
\title{Quantum interference terms in nonmesonic weak decay of $\Lambda$-hypernuclei
within a RPA formalism.}
\author{E. Bauer}

\affiliation{
Departamento de F\'{\i}sica, Universidad Nacional de
La Plata,\\
C. C. 67, 1900 La Plata, Argentina}
\affiliation{
Instituto de F\'{\i}sica La Plata,
CONICET, 1900 La Plata, Argentina}

\begin{abstract}
Single and double coincidence nucleon spectra in the
$\Lambda$-hypernuclei weak decay are evaluated and discussed using
a microscopic formalism. Nuclear matter is employed together with
the local density approximation which allows us to analyze the
$^{12}_{\Lambda}C$ hypernucleus non-mesonic weak decay. Final
state interactions ($FSI$) are included via the first order (in
the nuclear residual interaction) terms to the RPA, where the
strong residual interaction is modelled by a Bonn potential. At
this level of approximation, these $FSI$ are pure quantum
interference terms between the primary decay $(\Lambda N
\rightarrow NN)$ and $(\Lambda N \rightarrow NN \rightarrow NN)$,
where the strong interaction is responsible for the last piece in
the second reaction. Also the Pauli exchange contributions are
explicitly evaluated. We show that the inclusion of Pauli exchange
terms is important. A comparison with data is made. We conclude
that the limitations in phase space in the RPA makes this
approximation inadequate to reproduce the nucleon spectra. This
fact, does not allow us to draw a definite conclusion about the
importance of the interference terms.

\vspace{2.6cm}


\vspace{.5cm}

\end{abstract}

\maketitle

\newpage

\section{INTRODUCTION}
\label{INTR}
An hypernucleus is a bound system of nonstrange and strange baryons and
it is an important source of information
about baryon-baryon strangeness-changing weak interactions.
In the present contribution, we focus on the $\Lambda$-hypernuclei,
which can be produced by an hadronic reaction - such
as $(\pi, K)$ - or by an electromagnetic one - such as
$(e,e'K)$. The $\Lambda$ is generally formed in an excited state
and then, via the emission of a series of $\gamma$-rays,
it reaches its ground $(1s_{1/2})$-state. The $\Lambda$
decays itself mainly via the weak interaction by two
decay mechanisms: the so-called mesonic decay
$(\Lambda \rightarrow \pi N)$ and the non-mesonic
one $(NM)$, where no meson is present in the final state $(\Lambda N \rightarrow n N)$.
The mesonic decay is the main decay process for the free $\Lambda$, but
when the $\Lambda$ is within the nuclear environment it is
strongly inhibited by the Pauli principle .
In this case, the non-mesonic weak decay becomes the
most important decay mechanism. This decay can
be stimulated either by protons or neutrons, where
the corresponding decay widths are denoted as
$\Gamma_{p} \equiv \Gamma(\Lambda p \rightarrow n p)$ and
$\Gamma_{n} \equiv \Gamma(\Lambda n \rightarrow n n)$, respectively.
Just after the $\Lambda$-decays, the resultant nucleons are still within the
nucleus and in their way out they can interact strongly
with any nucleon of the nucleus.
Finally, two or more nucleons are ejected from the nucleus and it is
the nucleon spectra of these emitted nucleons the magnitude which
can be measured. For review articles one can see \cite{ra98,al02}, while some of the
experimental works are~\cite{mo74,ha01,sa05,ki02,ok04,bhang,outaVa,kang06,kim06}.

For the non-mesonic $\Lambda$-decay, there
are two quantities which deserves special attention.
The first one is the ratio $\Gamma_{n/p} \equiv \Gamma_n /
\Gamma_p$ where theory predicts a value smaller than 0.5, while
the so-called experimental result has a value closer to one. The second quantity is
the asymmetry of the protons emitted in the $NM$
decay of polarized hypernuclei. In this case, data indicate
a value close to zero, while most of the theoretical works predict a large negative number.
Just recently, the incorporation of the $\sigma$-meson suggests a solution to this
problem (see~\cite{ch07}, and references therein).
In the present work, however, we will not deal with the asymmetry. For $\Gamma_{n/p}$,
the connection between the theory and the experimental results is not straightforward because there
is no direct measurement of this ratio. Both $\Gamma_{n}$ and $\Gamma_{p}$
are the decay widths of the so-designed primaries disintegration and
as mentioned above, data give us results on the
the nucleon spectra which emerge from the nucleus. The connection
between $\Gamma_{n/p}$ and the nucleon spectra is a theoretical problem in
itself. Before going on, it is interesting to resume some of the theoretical
efforts in dealing with $\Gamma_{n/p}$.

The main ingredients required in
the evaluation of the transition rate $(\Lambda N \rightarrow n N)$,
are the transition potential and the wave functions which represent
the $\Lambda$ and the nucleons.
The first microscopic scheme for $\Gamma_{n/p}$ has been proposed by
Adams \cite{ad67}, who has used the  nuclear matter framework, one pion
exchange model (OPE), $\Delta T = 1/2$ piece of the $\Lambda N
\pi$ coupling and short range correlations (SRC). While this
model fairly reproduce the total $NM$ decay width $\Gamma_{NM}$,
it produce ratios smaller than 0.20. Some of the
improvements over Adams' model are: $i)$ the inclusion
of heavier mesons than the pion in the
$\Lambda N \rightarrow N N$--transition potential
\cite{mc84,ni93,mo94,pa95,du96,pa97,pa99,ji01,pa02,it02,ba03};
$ii)$ The inclusion of interaction terms that violates the isospin $\Delta T = 1/2$ rule
has been considered in \cite{mal94,go97,pa98}. Alternatively,  $iii)$ the
transition potential can be described in terms of the quark
degree of freedom~\cite{ch83,he86,ma95,ino94,sa00},
which automatically introduces the $\Delta T = 3/2$ contribution. And
$iv)$ the employment of finite nucleus wave functions instead of plane waves, a scheme usually called
Wave Function Method (WFM)~\cite{ni93,mo94,pa95,pa97,go97,pa98,pa99,pa02,it02}.
This list does not pretend to be complete. However, we should mention that in all
these works the discrepancy between theory and experiment remains.

We turn now to the interpretation of data. As mentioned, it is the
spectra of nucleons emerging from the nucleus, the quantity which is measured.
The nucleons originated in the $\Lambda$-decay interact strongly
with others nucleons before leaving the nucleus. These interactions are
called final state interactions $(FSI)$. Again, we can distinguish
two issues: $i)$ the primary decays and $ii)$ a model for the $FSI$.
Regarding the first point, we should note that
$(\Lambda N \rightarrow n N)$ is not the only $NM$-weak
decay mechanism: the $NM$-decay width can be
also stimulated by two nucleons $(\Lambda NN \rightarrow nNN)$.
The corresponding decay width is called $\Gamma_{2}$. This
two-body induced decay is originated from ground state correlation in the
hypernuclei \cite{al91,ra94,ra97,al00b,al00,ba04}. An estimation of
$\Gamma_{2}$ is important in the analysis of data, because this process is a source of
nucleons which can not be distinguished from those stemming from
$\Gamma_{1} \equiv \Gamma_{n} + \Gamma_{p}$.

There are several models for the $FSI$ among which we focus on
two of them: the intranuclear cascade code (INC)~\cite{ra97,ga03,ga04,ba06}
and the microscopic model developed in~\cite{ba07}.
The INC is a semi-phenomenological approach. The starting point
in the INC is the microscopic evaluation of
$\Gamma_{n}$, $\Gamma_{p}$ and $\Gamma_{2}$. Afterwards, the nucleons
produced in the weak decay are followed in a semi-classical manner until
they leave the nucleus. By means of this emulation of the physical
conditions of the hypernuclear decay, a more accurate agreement between
the theoretical results and the data is achieved. The analysis of
the experimental information using the INC, produce ratios
$(\Gamma_{n/p})^{exp} \sim 0.4-0.6$. However, one
limitation of the INC is that the quantum interference terms
between the primary weak decay reaction $(\Lambda N \rightarrow NN)$ and any
other reaction which has the same initial and final state, such as
$(\Lambda N \rightarrow NN \rightarrow NN)$, can not be included.
On the other hand and to the best of our knowledge, the microscopic model
described in~\cite{ba07}, is the only microscopic model which puts in the same
level of theoretical effort the weak decay mechanism and the $FSI$. Due to
its character, the microscopic model automatically includes the
quantum interference terms. Some details of the microscopic model
are given in the next Section.

While the theoretical prediction starts with the primary decay and should
end in the nucleon spectra, the analysis of the experiment begins with the spectra,
goes back and a value for $\Gamma_{n/p}$ is determined. This is done
using several models, among which the INC is certainly one of the
more elaborated ones. It should be noted, that the INC is employed
in two ways: for the theoretical prediction of the nucleon spectra
and for the interpretation of data.
In any case, for the extraction
of the $\Gamma_{n/p}$-experimental value one needs
two theoretical inputs: $\Gamma_{2}$ and the $FSI$.
Now, if several models for these inputs
produce the same ratio $\Gamma_{n/p}$, it is reasonable
to name it as the experimental value for $\Gamma_{n/p}$.
Therefore, an accurate evaluation of the quantum interference
terms together with an alternative formalism to the INC, is important.

In the present contribution, we further developed the
microscopic model of~\cite{ba07} presenting results for the
nucleon spectra, with emphasis in the quantum interference terms.
The paper is organized as follows. In Section~II we present
the microscopic model for the nucleon spectra. This is done
in general terms, including the two-body induced contribution.
In Section~III, explicit expression within the first
order contribution to the RPA are shown. In Section~IV, the
numerical results are discussed and finally, in Section~V we
give our conclusions.

\newpage
\section{THE MICROSCOPIC MODEL FOR THE NUCLEON SPECTRA}
\label{model}
In this Section the microscopic model developed in~\cite{ba07} is
briefly summarized with the addition of $\Gamma_{2}$. In fact,
$\Gamma_{2}$ is the sum of three terms:
$\Gamma_{nn} \equiv \Gamma(\Lambda nn \rightarrow nnn)$,
$\Gamma_{np} \equiv \Gamma(\Lambda np \rightarrow nnp)$ and
$\Gamma_{pp} \equiv \Gamma(\Lambda pp \rightarrow npp)$
(for details see~\cite{ba04}).
We are interested
in reporting expressions for the single and double nucleon spectra:
$N_{N}$ and $N_{NN'}$, which represent the number
of nucleons of kind $N=n, \,p$ vs. it kinetic
energy, $T_{N}$ and number of pairs of nucleons $NN'$ vs.
$T_{N}+T_{N'}$ or vs. the relative angle between $N$ and $N'$, $cos(\theta_{NN'})$,
respectively. Without $FSI$, the expressions for $N_{N}$ and $N_{NN'}$ are,
\begin{eqnarray}
\label{nn10}
N^{0}_{n} & = & 2 \bar{\Gamma}_{n} +  \bar{\Gamma}_{p} +
3 \bar{\Gamma}_{nn} + 2 \bar{\Gamma}_{np} + \bar{\Gamma}_{pp}\\
\label{np10}
N^{0}_{p} & = & \bar{\Gamma}_{p} + \bar{\Gamma}_{np} + 2
\bar{\Gamma}_{pp}, \\
\label{nnn10}
N^{0}_{nn} & = & \bar{\Gamma}_{n} + 3 \bar{\Gamma}_{nn}
+ \bar{\Gamma}_{np}\\
\label{nnp10}
N^{0}_{np} & = & \bar{\Gamma}_{p} + 2 \bar{\Gamma}_{np}
+ 2 \bar{\Gamma}_{pp} \\
\label{npp10}
N^{0}_{pp} & = & \bar{\Gamma}_{pp}
\end{eqnarray}
where we have used the normalization,
$\bar{\Gamma}_{N} \equiv \Gamma_{N}/\Gamma_{NM}$, with
$\Gamma_{NM}=\Gamma_{1}+\Gamma_{2}$.
The index $0$, indicates that there is no $FSI$. The multiplicative
factors in the right hand side of these equations are
the number of particles (pairs of particles)
of kind $N$ ($NN'$) produces by the primary decay. Explicit
expressions for $\Gamma_{1}$ and $\Gamma_{2}$ are
reported in~\cite{ba03} and~\cite{ba04}, respectively.
As mentioned, the nucleon spectra is the number of
particles (or pairs of particles) which angle or kinetic energy lays within
a certain range. Therefore, $\Gamma_{1}$ and $\Gamma_{2}$
are cut into pieces which correspond to certain
angles- or kinetic energies-ranges. This is easily implemented
by adding steps functions (in the integrand of $\Gamma_{1}$
and $\Gamma_{2}$), which limit the integration to such ranges.

The next step is the inclusion of the $FSI$. To this end,
we introduce the quantity $\Gamma_{i, i' \rightarrow j}$. This function
result from evaluating any possible Goldstone diagram for
the $\Lambda$-weak decay, where the strong interaction is present after
the weak decay takes place.
The index, $j$, is the final state (\textit{i.e.} the emitted nucleons),
taking the values: $j$=$nn$, $np$, $nnn$, $nnp$, etc.
At variance, the indices $i, \, i'$ refer to the two primary weak
decays of each diagram and can have the values $i \, ($or $i')=n, \, p,
\, nn, \, np, \, pp$; which stand for the transitions amplitudes $(\Lambda n \rightarrow nn)$,
$(\Lambda p \rightarrow np)$, $(\Lambda nn \rightarrow nnn)$, $(\Lambda np \rightarrow nnp)$
and $(\Lambda pp \rightarrow npp)$, respectively. It is important to be aware of the
fact that the decay widths,
$\Gamma_{n}$, $\Gamma_{p}$, $\Gamma_{nn}$, etc., are the square of a transition
amplitude ($(\Lambda N \rightarrow nN)$ or $(\Lambda NN \rightarrow nNN)$).
In this sense, $\Gamma_{i, i' \rightarrow j}$
does not represent only a decay width, but also some interference terms,
which are pure quantum mechanical effects.

By the addition of the $FSI$ in Eqs.~(\ref{nn10}-\ref{npp10}) we obtained,
\begin{eqnarray}
\label{nn1f}
N_{n} & = & 2 \bar{\Gamma}_{n} +  \bar{\Gamma}_{p} +
3 \bar{\Gamma}_{nn} + 2 \bar{\Gamma}_{np} + \bar{\Gamma}_{pp} +
\sum_{i, \, i'; \, j} N_{j \, (n)} \,
\bar{\Gamma}_{i, i' \rightarrow j},\\
\label{np1f}
N_{p} & = & \bar{\Gamma}_{p} + \bar{\Gamma}_{np} + 2
\bar{\Gamma}_{pp}, +
\sum_{i, \, i'; \, j} N_{j \, (p)} \, \bar{\Gamma}_{i, i' \rightarrow j},\\
\label{nnn1f}
N_{nn} & = & \bar{\Gamma}_{n} + 3 \bar{\Gamma}_{nn}
+ \bar{\Gamma}_{np}+
\sum_{i, \, i'; \, j} N_{j \, (nn)} \, \bar{\Gamma}_{i, i' \rightarrow j},\\
\label{nnp1f}
N_{np} & = & \bar{\Gamma}_{p} + 2 \bar{\Gamma}_{np}
+ 2 \bar{\Gamma}_{pp} +
\sum_{i, \, i'; \, j} N_{j \, (np)} \, \bar{\Gamma}_{i, i' \rightarrow j},\\
\label{npp1f}
N_{pp} & = & \bar{\Gamma}_{pp} +
\sum_{i, \, i'; \, j} N_{j \, (pp)} \, \bar{\Gamma}_{i, i' \rightarrow j},
\end{eqnarray}
where the factors $N_{j \, (N)}$ are the numbers of nucleons of the type $N$
in the state $j$. In the same way, $N_{j \, (NN')}$ are the numbers of
pairs of nucleons of the type $NN'$ in the state $j$. Let us give two
examples. If $j=np$, $N_{np \, (n)}=1$, $N_{np \, (p)}=1$, $N_{np \, (nn)}=0$,
$N_{np \, (np)}=1$, $N_{np \, (pp)}=0$; and if
$j=nnn$, $N_{nnn \, (n)}=3$, $N_{nnn \, (p)}=0$, $N_{nnn \, (nn)}=3$,
$N_{nnn \, (np)}=0$, $N_{nnn \, (pp)}=0$. In these expressions, the
summation over $i, \, i'$ and $j$, runs over the values of these
indices mentioned above.

As a final comment for this section we make some further considerations
about the quantum interference terms. All decay widths come from
the square of a transition amplitude. When no $FSI$ are included, the only
transitions amplitudes are $(\Lambda N \rightarrow nN)$ and
$(\Lambda NN \rightarrow nNN)$. For simplicity, we focus on
$(\Lambda N \rightarrow nN)$. There is no interference term
between $(\Lambda n \rightarrow nn)$ and $(\Lambda p \rightarrow np)$
because the final state is different.
When the $FSI$ come into play, the strong interaction allows
many others transitions amplitudes. In this case, as different
reactions can end in the same final state and as the initial
state is the same hypernuclei state for all the processes, the total transition
amplitude is the sum of all these terms. The squares of the
individual terms, are the decay
widths, while the crossed products are the interference terms.
Within our model, $\Gamma_{i, i' \rightarrow j}$ contain both
decay widths and interference terms. Note that the strong interaction is present
in $\Gamma_{2}$ as a ground state correlation, that is, the
strong interaction acts before the weak transition potential.
As an additional comment about the interference terms,
they can be grouped into two categories: $a)$ the ones
with $i = i'$ and $b)$ the ones with $i \neq i'$.
It is perhaps more convenient to explain these categories
by means of an example. Let us propose three transition amplitudes:
$A_{1}: \, \Lambda n \rightarrow nn$,
$A_{2}: \, \Lambda n \rightarrow nn \rightarrow nn$ and
$A_{3}: \, \Lambda p \rightarrow np \rightarrow nn$, where
in the last two expressions the strong interaction is responsible
for the second reaction. There is an interference term between
$A_{1}$ and $A_{2}$ ($i = i'$), where the strong interaction
appears in first order. It is clear that the square of
$A_{1}+A_{2}$ is positive, but the interference terms between
both terms can be either positive or negative. Finally, the interference
terms between $A_{1}$ and $A_{3}$ is an example of the
$i \neq i'$-category.

\newpage
\section{EXPLICIT EXPRESSIONS FOR $\Gamma_{i, i' \rightarrow j}$}
\label{rpa}

In this section we present expressions for the
functions $\Gamma_{i, i' \rightarrow j}$, using
non-relativistic nuclear matter together with the
Local Density Approximation (LDA)~\cite{os85}, which
allows us to discuss any particular hypernucleus.
The Eqs.~(\ref{nn1f}-\ref{npp1f}) are general expressions for
$N_{N}$ and $N_{NN'}$. Although $\Gamma_{i, i' \rightarrow j}$ is
completely defined, one has to choose some set of Goldstone diagrams
which represent the $\Lambda$-decay to obtain the explicit
expressions for this function. This set has an
infinite number of diagrams and \textit{a priori} any sub-set of
diagrams could be equally important. In this work and as a first
step in the evaluation of the nucleon spectra, we have decided
to study the RPA-like diagrams. The direct part of the RPA, known
as ring approximation, is perhaps the simplest manner
to implement the strong interaction into this problem.
The Pauli exchange terms are known to be important
for these particular set of diagrams (see~\cite{ba96}) and
consequently the RPA looks as a natural first step.
In~\cite{ba07} we have presented expressions for
$\Gamma_{i, i' \rightarrow j}$ within the ring approximation.
However, no comparison with data has been done, because of
the absence of the Pauli exclusion principle.

Let us call by $\Gamma$ a
generic function which can be either $\Gamma_{1}$, $\Gamma_{2}$
or $\Gamma_{i, i' \rightarrow j}$.
Instead of giving the expression for $\Gamma$, it is more
convenient to work with the partial decay width
$\Gamma(k,k_{F_{n}}, k_{F_{p}})$,
where, $k$ is the $\Lambda$ energy-momentum,
$k_{F_{n}}$ and $k_{F_{p}}$ are the Fermi momentum
for neutrons and protons, respectively.
To evaluate $\Gamma(k)$ for a particular nucleus one can use
either an effective Fermi momentum or the Local Density
Approximation. In this work the LDA is adopted, which make
$k_{F_{n}}$ and $k_{F_{p}}$ position-dependent. They are defined as
$k_{F_{n \, (p)}}(r) = \hbar c (3 \pi^{2} \rho_{n \, (p)}(r))^{1/3}$, where
$\rho_n(r)=\rho(r) N/(N+Z)$ and $\rho_p(r)=\rho(r) Z/(N+Z)$, with
$\rho(r)$, $N$ and $Z$ being, respectively, the density profile,
number of neutrons and number of protons of the nuclear core of the hypernucleus.
In the last case,
it is equivalent to write the function $\Gamma(k,k_{F_{n}}, k_{F_{p}})$
in terms of the densities as $\Gamma(k,\rho_n(r), \rho_p(r))$.
The LDA reads,
\begin{equation}
\label{decwpar3}
\Gamma(k) = \int d \v{r} \,
\Gamma(k,\rho_n(r), \rho_p(r)) \;  |\psi_{\Lambda}(\v{r})|^2 \,
\end{equation}
where for the $\Lambda$ wave function $\psi_{\Lambda}(\v{r})$, we
take the $1s_{1/2}$ wave function of a harmonic oscillator.
The final result is obtained by averaging over the $\Lambda$ momentum
distribution, $|\widetilde{\psi}_{\Lambda}(\v{k})|^2$, as follows,
\begin{equation}
\label{decwpar2}
\Gamma = \int d \v{k} \,
\Gamma(k) \;  |\widetilde{\psi}_{\Lambda}(\v{k})|^2 \,
\end{equation}
where $\widetilde{\psi}_{\Lambda}(\v{k})$ is the Fourier transform of
$\psi_{\Lambda}(\v{r})$ and
$k_{0}=E_{\Lambda}(\v{k})+V_{\Lambda}$, being $V_{\Lambda}$
the binding energy for the $\Lambda$.

\begin{figure}[h]
\centerline{\includegraphics[scale=0.57]{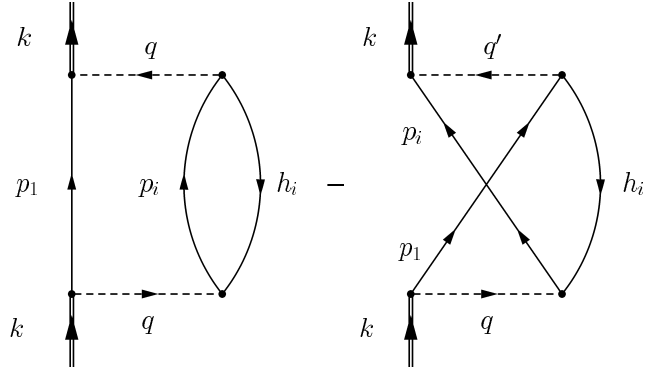}}
\caption{Goldstone diagrams for the
direct and exchange contributions
to the $\Lambda$ decay width, respectively.}
\label{fig1rpa}
\end{figure}

As remarked in the last section, the expressions
for $\Gamma_{i, i' \rightarrow j}$ are obtained as if they
were decay widths. In this spirit, it is convenient to overview
the derivation of $\Gamma_{1}$, because this simplifies the derivation
of $\Gamma_{i, i' \rightarrow j}$.
In Fig.~\ref{fig1rpa} we show the direct and exchange $\Gamma_{1}$-contributions.
The distinction between $\Gamma_{n}$ and $\Gamma_{p}$ can be
ascribed to the isospin of the hole line, where when $h_{i}$
is a neutron (proton)-hole we are considering the
$\Gamma_{n}$ ($\Gamma_{p}$)-decay width.
To get the analytical expressions for
these diagrams we have employed the Golstone rules
(for a more detailed version, see~\cite{ba03}),
\begin{eqnarray}
\label{gamdirpi0}
\Gamma_{t_{hi}}^{ant}(k,k_{F_{n}}, k_{F_{p}})  & =
& -2 \, Im \, \int \frac{d^4 \, p_{1}}{(2 \pi)^4} \,
\int \frac{d^4 \, p_{i}}{(2 \pi)^4} \;  G_{part}(p_{1}) \; G_{part}(p_{i})
\; G_{hole}(h_{i}) \, \frac{1}{4}
\nonumber \\
&& \sum_{\mbox{\tiny all spins}, \; t_{p 1}, \, t_{pi}} \;
\bra{\gamma_{\Lambda}}
(V^{\Lambda N})^{\dag}
\ket{\gamma_{p1} \gamma_{pi} \gamma_{hi}}_{ant} \;
\bra{\gamma_{p1} \gamma_{pi} \gamma_{hi} }
V^{\Lambda N}
\ket{\gamma_{\Lambda}}_{ant},
\end{eqnarray}
where for simplicity,
$\gamma_{i}$ represents the spin ($s$), isospin ($t$) and
energy-momentum of the particle $i$.
The energy-momentum carried by the transition potential,
$V^{\Lambda N}$, can not be specified until we separate the
direct and the exchange terms. The meaning of the subindexes $p i$ and $h i$
is shown in Fig.~\ref{fig1rpa}, where due to the momentum conservation,
$\v{h_{i}}=\v{p_{1}}+\v{p_{i}}-\v{k}$ . The direct and exchange
matrix elements are,
\begin{eqnarray}
\label{me0ant}
\bra{\gamma_{\Lambda}}
(V^{\Lambda N})^{\dag}
\ket{\gamma_{p1} \gamma_{pi} \gamma_{hi}}_{ant} \;
\bra{\gamma_{p1} \gamma_{pi} \gamma_{hi} }
V^{\Lambda N}
\ket{\gamma_{\Lambda}}_{ant} = \nonumber \\
= \bra{\gamma_{\Lambda}}
(V^{\Lambda N}(q))^{\dag}
\ket{\gamma_{p1} \gamma_{pi} \gamma_{hi}}
 \bra{\gamma_{p1} \gamma_{pi} \gamma_{hi} }
V^{\Lambda N}(q)
\ket{\gamma_{\Lambda}} \, - \nonumber \\
- \, \bra{\gamma_{\Lambda}}
(V^{\Lambda N}(q'))^{\dag}
\ket{\gamma_{pi} \gamma_{p1} \gamma_{hi}}
 \bra{\gamma_{p1} \gamma_{pi} \gamma_{hi} }
V^{\Lambda N}(q)
\ket{\gamma_{\Lambda}}
\end{eqnarray}
where the minus sign
comes from  the crossing  of the fermionic lines.
Due to energy-momentum conservation, we have
$q = k - p_{1}$ and $q' = k - p_{2}$.
The first (second) term in the the right hand side
of Eq.~(\ref{me0ant}) originates the direct (exchange)
contribution to the decay width.

The particle and hole propagators are,
\begin{equation}
\label{nprop}
G_{part}(p)  =  \frac{\theta(|\mbox{\boldmath $p$}| - k_F)}
{p_0 - E_N(\mbox{\boldmath $p$})- V_N + i
\varepsilon}
\end{equation}
and
\begin{equation}
\label{nprop2}
G_{hole}(h)  = \frac{\theta(k_F - |\mbox{\boldmath $h$}| )}
{h_0 - E_N(\mbox{\boldmath $h$})- V_N - i
\varepsilon},
\end{equation}
where $E_N$ is the nucleon
total free energy and $V_N$ is the nucleon binding energy.
The value of the Fermi momentum depends on whether we have
a proton or a neutron.
The $\Lambda N \rightarrow NN$ transition
potential $V^{\Lambda N}$, is,
\begin{equation}
\label{intlnnn}
V^{\Lambda N (NN)} (q) = \sum_{\tau_{\Lambda (N)}=0,1}
 {\cal O}_{\tau_{\Lambda (N)}}
{\cal V}_{\tau_{\Lambda (N)}}^{\Lambda N (NN)} (q),
\end{equation}
where we have included the
nuclear residual interaction $V^{NN}$, which
is employed soon. The isospin dependence is given by,
\begin{eqnarray}
\label{isos} {\cal O}_{\tau_{\Lambda (N)}} =~~~~~
\left\{
\begin{array}{c}1,~~\mbox{for}~~\tau_{\Lambda (N)}=0\\
  \v{\tau}_1 \cdot \v{\tau}_2,~~\mbox{for}~~\tau_{\Lambda (N)}=1
\end{array}\right.
\end{eqnarray}
The values $\tau=0,1$ stand for the isoscalar
and isovector parts of the interaction, respectively.
The spin and momentum dependence of the transition potential is,
\begin{eqnarray}
\label{intln}
{\cal V}_{\tau_{\Lambda}}^{\Lambda N} (q) &
= &  (G_F m_{\pi}^2)  \; \{
S_{\tau_{\Lambda}}(q)  \; \v{\sigma}_1 \cdot \v{\hat{q}} +
S'_{\tau_{\Lambda}}(q)  \; \v{\sigma}_2 \cdot \v{\hat{q}} +
P_{L, \tau_{\Lambda}}(q)
 \v{\sigma}_1 \cdot \v{\hat{q}} \; \v{\sigma}_2 \cdot
\v{\hat{q}} + P_{C, \tau_{\Lambda}}(q)  +  \nonumber \\
& & +  P_{T, \tau_{\Lambda}}(q)  (\v{\sigma}_1 \times \v{\hat{q}})
\cdot  (\v{\sigma}_2 \times \v{\hat{q}}) + i S_{V, \tau_{\Lambda}}(q)
\v{(\sigma}_1 \times \v{\sigma}_2) \cdot
\v{\hat{q}} \},
\end{eqnarray}
where the quantities  $S_{\tau_{\Lambda}}(q)$, $S'_{\tau_{\Lambda}}(q)$,
$P_{L, \tau_{\Lambda}}(q)$, $P_{C,\tau_{\Lambda}}(q)$,
$P_{T, \tau_{\Lambda}}(q)$ and $S_{V, \tau_{\Lambda}}(q)$ contain
short range correlations (SRC) and  are given in Appendix B of \cite{ba03}.
They are built up from the full one-meson-exchange potential (OMEP),
which involves the  complete pseudoscalar and vector meson octets
($\pi,\eta,K,\rho,\omega,K^*$).
It is self evident that the $S$ ($P$)-terms are the parity violating (parity conserving)
terms of the transition potential.

The nuclear residual interaction is drawn as,
\begin{equation}
\label{intnn}
{\cal V}_{\tau_N}^{N N} (q)
= (\frac{f_{\pi}^2}{m_{\pi}^2})  \; \{
{\cal V}_{C, \,\tau_{N}}(q) +
{\cal V}_{\sigma, \, \tau_{N}}(q)
\v{\sigma}_1 \cdot \v{\sigma}_2 +
{\cal V}_{L, \, \tau_{N}}(q)
\v{\sigma}_1 \cdot \v{\hat{q}} \; \v{\sigma}_2 \cdot
\v{\hat{q}} \}.
\end{equation}
where the functions ${\cal V}_{C, \,\tau_{N}}(q)$,
${\cal V}_{\sigma, \, \tau_{N}}(q)$ and
${\cal V}_{L, \, \tau_{N}}(q)$ are adjusted to reproduce
any effective OMEP-nuclear residual interaction.

\begin{figure}[h]
\centerline{\includegraphics[scale=0.75]{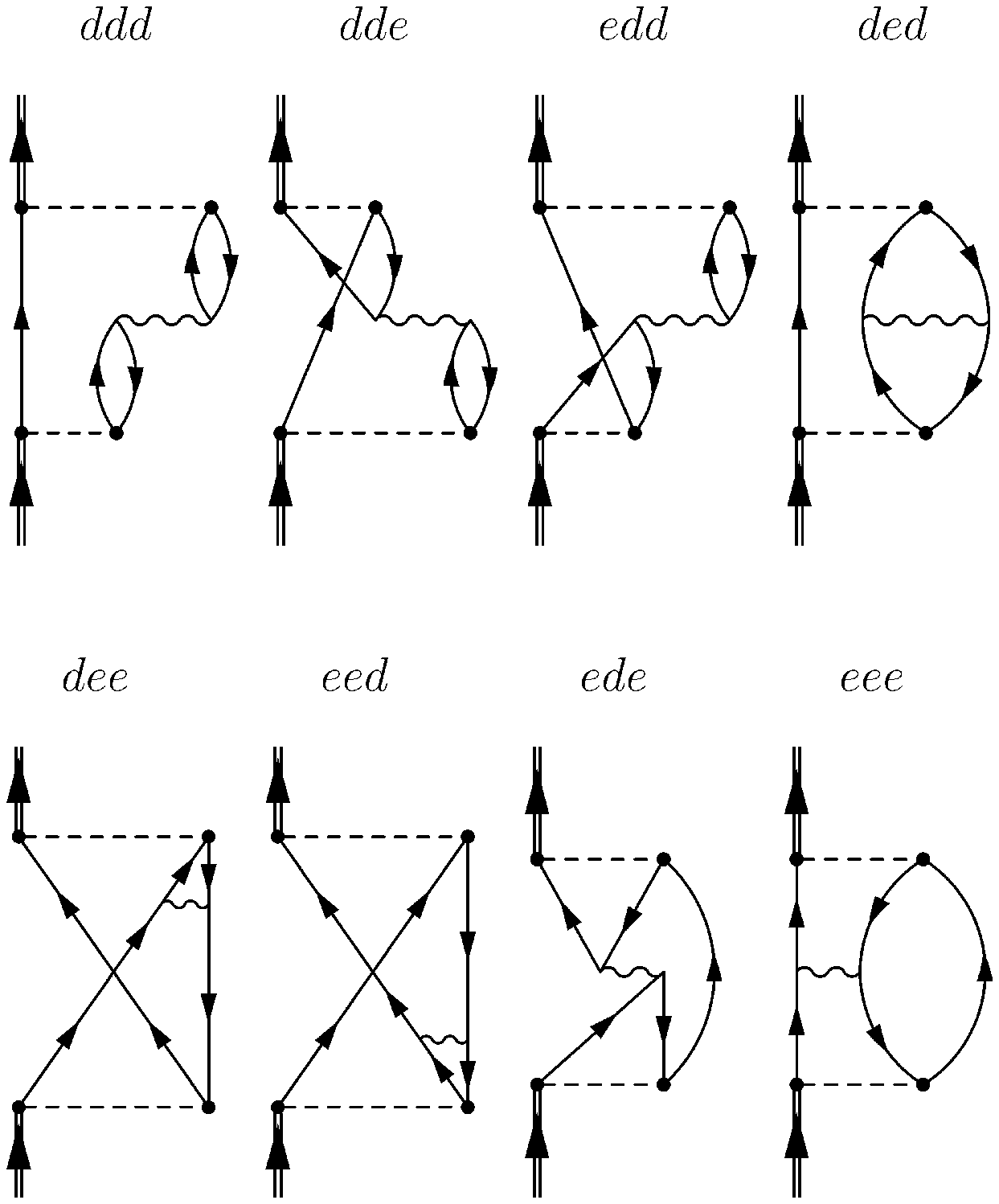}}
\caption{Goldstone diagrams for the first order contribution to
the RPA-series. The meaning of the notation is explained in the
text. The dashed and wavy lines stand for $V^{\Lambda N}$ and
$V^{NN}$, respectively. An up (down) going arrow represents a
particle (hole), while an arrow with a wide line represents the
$\Lambda$.}
\label{fig2rpa}
\end{figure}

\begin{figure}[h]
\centerline{\includegraphics[scale=0.7]{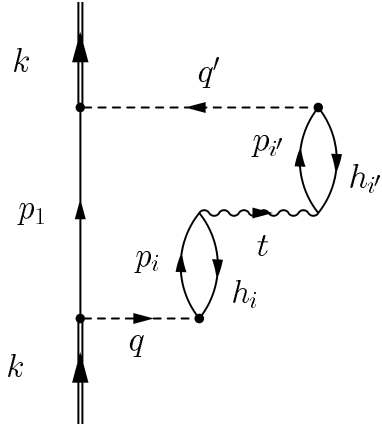}}
\caption{Here we have repeated the $ddd$-contribution
from Fig.~\ref{fig3rpa}, in order to assign values to the momentum
transfer. The remainder contributions have the same pattern.}
\label{fig3rpa}
\end{figure}

We have now all the elements required for the evaluation
of $\Gamma_{i, i' \rightarrow j}$ within the RPA approximation.
The ring approximation has
been discussed in~\cite{ba07}, where the advantage of using that approximation
is that the ring series can be summed up to infinite order
in a very simple way. The situation is different in the
RPA, where the corresponding series can be summed up only in
some particular cases: when the nuclear residual interaction is
represented by a contact or by a separable interaction. For a general finite range interaction
there is no way out but to evaluate each exchange
term individually. The problem is quite involved because
matrix elements must be antisymmetrized for both
$V^{\Lambda N}$ and $V^{NN}$. The lowest order RPA-contribution
is the one in which $V^{NN}$ appears in first order, being
the only one reported in the present contribution.
In the lowest RPA-contribution,
there are two matrix elements with $V^{\Lambda N}$ and one
with $V^{NN}$. Each matrix element has a direct and an exchange
part. This makes a total of eight different diagrams, which
are shown in Fig.~\ref{fig2rpa}. Unfortunately, in nuclear matter each
diagram must be evaluated individually.
In the next section and by means of the numerical analysis, we discuss
this approximation. Using the standard Goldstone rules,
this lowest order RPA-contribution is written as,
\begin{eqnarray}
\label{gamgold}
\Gamma_{i, \, i' \rightarrow j \;}(k,k_{F_{n}}, k_{F_{p}}) & = &
 -2 \, Im \, \int \frac{d^4 \, p_{1}}{(2 \pi)^4} \,
\int \frac{d^4 \, h_{i}}{(2 \pi)^4} \, \, \int \frac{d^4 \, h_{i'}}{(2 \pi)^4} \;
\frac{1}{4} \sum
G_{part}(p_{1}) G_{part}(p_{i})
\nonumber \\
&&
G_{part}(p_{i'}) G_{hole}(h_{i}) G_{hole}(h_{i'})
\nonumber \\
&& \times \bra{\gamma_{\Lambda}}
(V^{\Lambda N}(q'))^{\dag}
\ket{\gamma_{p1} \gamma_{pi'} \gamma_{hi'}}_{ant}
\bra{\gamma_{p1} \gamma_{pi'} \gamma_{hi'}} V^{N N}(t)
\ket{\gamma_{p1} \gamma_{pi} \gamma_{hi}}_{ant}
\nonumber \\
&& \times \bra{\gamma_{p1} \gamma_{pi} \gamma_{hi}}
V^{\Lambda N}(q)
\ket{\gamma_{\Lambda}}_{ant}.
\end{eqnarray}
The summation runs over all spins, while the isospin
sum can not be specified until one sets the final state $j$.
The energy-momentum carried by each fermionic line is shown in
Fig.~\ref{fig3rpa} for the direct term.
Due to the energy-momentum conservation in each vertex,
for all the contributions we have
$p_{i}=h_{i}+k-p_{1}$ and $p_{i'}=h_{i'}+k-p_{1}$; while the
energy-momentum carried by $V^{N N}$ and $V^{\Lambda N}$ depends on the
topology of each diagram. These values, together with some
more details on the exchange terms are specified in the Appendix.

It is convenient to re-write Eq.~(\ref{gamgold}) as,
\begin{equation}
\label{direxch}
\Gamma_{i, \, i' \rightarrow j \;} = \sum_{\alpha, \, \beta, \, \delta=d,e} \;
\Gamma^{\alpha \beta \delta }_{i, \, i' \rightarrow j \;},
\end{equation}
where the super-indexes $\alpha, \, \beta$ and  $\delta$, refer
to the direct ($d$) and exchange ($e$) matrix elements
$\bra{}(V^{\Lambda N})^{\dag}\ket{}_{d+e}$,
$\bra{} V^{N N} \ket{}_{d+e}$ and
$\bra{} V^{\Lambda N} \ket{}_{d+e}$, respectively.
For instance, for the direct
contribution, we have $\alpha \, \beta \, \delta =\, ddd$.
Note that $\alpha$ ($\beta$ and $\delta$), can take
the 'value' $d$ or $e$, which makes the total of eight
different contributions, being $ddd$ the
first order term to the ring series.

By inspection of the diagrams in Fig.~\ref{fig2rpa}, we notice that
the strong interaction split these graphs into two pieces. In the
lower piece particles are named as $p_{1}$ and $p_{i}$,
while in the upper portion we have $p_{1}$ and $p_{i'}$.
Each diagram gives two contributions: the first one with
$p_{1}$ and $p_{i}$ ($p_{1}$ and $p_{i'}$) on (off) the
mass shell, and the inverse situation.
Then, it is convenient to differentiate
these two possibilities by the super-index $l$ and $u$, respectively, as
follows,
\begin{eqnarray}
\label{rpa3b}
\Gamma^{\alpha  \beta  \delta , \, u}_{i, \, i' \rightarrow j \;}(k,k_{F_{n}}, k_{F_{p}}) &
\equiv &
\Gamma^{\alpha  \beta  \delta}_{i, \, i' \rightarrow j \;}(k,k_{F_{n}}, k_{F_{p}})|_{
p1, \, pi' \, on \, the \, mass \, shell}
\nonumber \\
\Gamma^{\alpha  \beta  \delta , \, l}_{i, \, i' \rightarrow j \;}(k,k_{F_{n}}, k_{F_{p}}) &
\equiv &
\Gamma^{\alpha  \beta  \delta}_{i, \, i' \rightarrow j \;}(k,k_{F_{n}}, k_{F_{p}})|_{
p1, \, pi \, on \, the \, mass \, shell}
\end{eqnarray}
The origin of these two terms comes from the energy integration
in Eq.~(\ref{gamgold}), which produces an expression with two
poles: one in the upper and the other in the lower part of the diagram.
Clearly, the distinction between $l$ and $u$, is
redundant for the direct contribution, due to the symmetry of the
diagram.

In order to perform the summation on spin and isospin quantum numbers
it is convenient to write,
\begin{eqnarray}
\label{rpa2}
\Gamma^{\alpha  \beta  \delta, \, l (u) }_{i, \, i' \rightarrow j \;}(k,k_{F_{n}}, k_{F_{p}}) & =
&\sum_{\tau, \tau_{N}, \tau'=0,1}
{\cal T}^{\alpha  \beta  \delta, \, l (u) }_{i, \, i' \rightarrow j; \; \tau' \tau_{N} \tau} \; \;
\widetilde{\Gamma}_{i, \, i' \rightarrow j; \;
\tau' \, \tau_{N} \, \tau}^{\alpha  \beta  \delta, \, l (u) }(k,k_{F_{n}}, k_{F_{p}})
\end{eqnarray}
where
\begin{eqnarray}
\label{rpa3}
{\cal T}^{\alpha  \beta  \delta, \, u}_{i, \, i' \rightarrow j; \; \tau' \tau_{N} \tau}
& = & \sum_{u} \, \bra{t_\Lambda} {\cal O}_{\tau'} \ket{t_{p1} t_{pi'} t_{hi'}}_{\alpha}
\bra{t_{pi'} t_{hi'}} {\cal O}_{\tau_{N}} \ket{t_{pi} t_{hi}}_{\beta}
\bra{t_{p1} t_{pi} t_{hi}} {\cal O}_{\tau} \ket{t_\Lambda}_{\delta} \nonumber \\
{\cal T}^{\alpha  \beta  \delta, \, l}_{i, \, i' \rightarrow j; \; \tau' \tau_{N} \tau}
& = & \sum_{l} \, \bra{t_\Lambda} {\cal O}_{\tau'} \ket{t_{p1} t_{pi'} t_{hi'}}_{\alpha}
\bra{t_{pi'} t_{hi'}} {\cal O}_{\tau_{N}} \ket{t_{pi} t_{hi}}_{\beta}
\bra{t_{p1} t_{pi} t_{hi}} {\cal O}_{\tau} \ket{t_\Lambda}_{\delta}
\end{eqnarray}
In the isospin summation, $l$ ($u$) means that the final state $j$ is
in the lower (upper) piece of the diagram. To avoid confusion, we
give an example: let $i=n$, $i'=p$ and $j=np$. The $l$ sum is zero because
for the final state $j$, there are no protons in the lower portion of the
diagram. The $u$ sum reduces to one term which isospin projections for
each particles are $t_{p1}=t_{pi}=t_{hi'}=-1/2$ and $t_{pi'}=t_{hi}=1/2$.

Performing the energy integration, the spin summation
and after some algebra, we obtain,
\begin{eqnarray}
\label{rpa4}
\widetilde{\Gamma}_{i, \, i' \rightarrow j; \,
\tau' \, \tau_{N} \, \tau}^{\alpha  \beta  \delta, \, l}  & =
& -(G_F m_{\pi}^2)^2 \frac{(-1)^{N_{F}}}{(2 \pi)^7}
(\frac{f_{\pi}^2}{m_{\pi}^2}) \, \int  \int \int
d \v{p}_{1} d \v{h}_{i} d \v{h}_{i'} \;
{\cal S}^{\alpha  \beta  \delta}_{\tau' \tau_{N} \tau}(q, q', t)  \;
\theta(q_0) \theta(q'_0)  \nonumber \\
& & \times \theta(|\v{p}_{1}|-k_{Fp1}) \theta(|\v{p}_{i}|-k_{Fpi}) \theta(k_{Fhi}-|\v{h}_{i}|)
\theta(|\v{p}_{i'}|-k_{Fpi'}) \theta(k_{Fhi'}-|\v{h}_{i'}|) \nonumber \\
& & \times \frac{ {\cal P} }{q'_0 - (E_N(\v{p}_{i'}) - E_N( \v{h}_{i'}))} \;
\delta(q_0 - (E_N(\v{p}_{i}) - E_N( \v{h}_{i})))
\\
\label{rpa4p}
\widetilde{\Gamma}_{i, \, i' \rightarrow j; \,
\tau' \, \tau_{N} \, \tau}^{\alpha  \beta  \delta, \, u}  & =
& -(G_F m_{\pi}^2)^2 \frac{(-1)^{N_{F}}}{(2 \pi)^7}
(\frac{f_{\pi}^2}{m_{\pi}^2}) \, \int  \int \int
d \v{p}_{1} d \v{h}_{i} d \v{h}_{i'} \;
{\cal S}^{\alpha  \beta  \delta}_{\tau' \tau_{N} \tau}(q, q', t)  \;
\theta(q_0) \theta(q'_0)  \nonumber \\
& & \times \theta(|\v{p}_{1}|-k_{Fp1}) \theta(|\v{p}_{i}|-k_{Fpi}) \theta(k_{Fhi}-|\v{h}_{i}|)
\theta(|\v{p}_{i'}|-k_{Fpi'}) \theta(k_{Fhi'}-|\v{h}_{i'}|) \nonumber \\
& & \times \frac{ {\cal P} }{q_0 - (E_N(\v{p}_{i}) - E_N( \v{h}_{i}))} \;
\delta(q'_0 - (E_N(\v{p}_{i'}) - E_N( \v{h}_{i'})))
\end{eqnarray}
In these expressions, $N_{F}$ is the number of crossing  of fermionic lines
and ${\cal P}$ indicates the principal value.
Explicit expressions for the
functions ${\cal S}^{\alpha  \beta  \delta}_{\tau' \tau_{N} \tau}(q, q', t)$
can be found in the Appendix. Here $q$ ($q'$) is the energy-momentum
carried by the transition potential in the lower (upper) part of the
diagram, while $t$ corresponds to the nuclear interaction.
We present in this section explicit expressions for the direct
contribution, whereas the exchange ones are shown in the Appendix.
The ${\cal S}^{ddd}_{\tau' \tau_{N} \tau}(q, q', t)$ reads,
\begin{eqnarray}
\label{tdir3}
{\cal S}^{ddd}_{\tau' \, \tau_{N} \, \tau}(q) & = &
4 \; \{ (S'_{\tau'} S'_{\tau} + P_{C, \tau'} P_{C, \tau})
{\cal V}_{C, \,\tau_{N}} + (S_{\tau'}
S_{\tau}
 + P_{L, \tau'}  P_{L, \tau}) {\cal V}_{L, \, \tau_{N}}
  + \nonumber \\
 & & + 2  \, ( S_{V, \tau'} S_{V, \tau} +
P_{T, \tau'} P_{T, \tau}) {\cal V}_{T, \, \tau_{N}} \}
\end{eqnarray}
where for the direct contribution we have, $q'=t=q$.
By means of Eqs.~(\ref{decwpar3}) and (\ref{decwpar2}), the
dependence on the Fermi and $\Lambda$-momenta in
the partial widths is eliminated.
For the $(ddd)$-diagrams, we have
$\widetilde{\Gamma}_{i, \, i' \rightarrow j; \, \tau' \, \tau_{N} \, \tau}^{ddd, \, l}=
\widetilde{\Gamma}_{i, \, i' \rightarrow j; \, \tau' \, \tau_{N} \, \tau}^{ddd, \, u} \equiv
\widetilde{\Gamma}_{i, \, i' \rightarrow j; \, \tau' \, \tau_{N} \, \tau}^{ddd}/2$.
By performing the summation over isospin for all possible primary decays and
final states, we have,
\begin{eqnarray}
\label{decnp2}
\Gamma^{ddd}_{n, n \rightarrow nn} & = &
                \widetilde{\Gamma}^{ddd, \, n}_{n, n \rightarrow nn, \, 111} +
                \widetilde{\Gamma}^{ddd, \, n}_{n, n \rightarrow nn, \, 000} +
                \widetilde{\Gamma}^{ddd, \, n}_{n, n \rightarrow nn, \, 110} +
                \widetilde{\Gamma}^{ddd, \, n}_{n, n \rightarrow nn, \, 101} +
                \widetilde{\Gamma}^{ddd, \, n}_{n, n \rightarrow nn, \, 011} +
\nonumber \\
          &&
              + \widetilde{\Gamma}^{ddd, \, n}_{n, n \rightarrow nn, \, 100} +
                \widetilde{\Gamma}^{ddd, \, n}_{n, n \rightarrow nn, \, 010} +
                \widetilde{\Gamma}^{ddd, \, n}_{n, n \rightarrow nn, \, 001}
\nonumber \\
\Gamma^{ddd}_{n, p \rightarrow nn} & = &
                \widetilde{\Gamma}^{ddd, \, n}_{n, p \rightarrow nn, \, 111} +
                \widetilde{\Gamma}^{ddd, \, n}_{n, p \rightarrow nn, \, 000} +
                \widetilde{\Gamma}^{ddd, \, n}_{n, p \rightarrow nn, \, 110} -
                \widetilde{\Gamma}^{ddd, \, n}_{n, p \rightarrow nn, \, 101} -
                \widetilde{\Gamma}^{ddd, \, n}_{n, p \rightarrow nn, \, 011} -
\nonumber \\
          &&  - \widetilde{\Gamma}^{ddd, \, n}_{n, p \rightarrow nn, \, 100} -
                \widetilde{\Gamma}^{ddd, \, n}_{n, p \rightarrow nn, \, 010} +
                \widetilde{\Gamma}^{ddd, \, n}_{n, p \rightarrow nn, \, 001}
\nonumber \\
\Gamma^{ddd}_{p, n \rightarrow nn} & = &
                \widetilde{\Gamma}^{ddd, \, n}_{p, n \rightarrow nn, \, 111}
              +  \widetilde{\Gamma}^{ddd, \, n}_{p, n \rightarrow nn, \, 000}
              -  \widetilde{\Gamma}^{ddd, \, n}_{p, n \rightarrow nn, \, 110}
              -  \widetilde{\Gamma}^{ddd, \, n}_{p, n \rightarrow nn, \, 101}
              +  \widetilde{\Gamma}^{ddd, \, n}_{p, n \rightarrow nn, \, 011} +
\nonumber \\
          &&  +  \widetilde{\Gamma}^{ddd, \, n}_{p, n \rightarrow nn, \, 100}
              -  \widetilde{\Gamma}^{ddd, \, n}_{p, n \rightarrow nn, \, 010}
              -  \widetilde{\Gamma}^{ddd, \, n}_{p, n \rightarrow nn, \, 001}
\nonumber \\
\Gamma^{ddd}_{n, p \rightarrow np} & = &
                \widetilde{\Gamma}^{ddd, \, n}_{n, p \rightarrow np, \, 111} +
                \widetilde{\Gamma}^{ddd, \, n}_{n, p \rightarrow np, \, 000} +
                \widetilde{\Gamma}^{ddd, \, n}_{n, p \rightarrow np, \, 110} -
                \widetilde{\Gamma}^{ddd, \, n}_{n, p \rightarrow np, \, 101} -
                \widetilde{\Gamma}^{ddd, \, n}_{n, p \rightarrow np, \, 011} -
\nonumber \\
          &&   -\widetilde{\Gamma}^{ddd, \, n}_{n, p \rightarrow np, \, 100} -
                \widetilde{\Gamma}^{ddd, \, n}_{n, p \rightarrow np, \, 010} +
                \widetilde{\Gamma}^{ddd, \, n}_{n, p \rightarrow np, \, 001}
\nonumber \\
\Gamma^{ddd}_{p, n \rightarrow np} & = &
                 \widetilde{\Gamma}^{ddd, \, n}_{p, n \rightarrow np, \, 111}
               + \widetilde{\Gamma}^{ddd, \, n}_{p, n \rightarrow np, \, 000}
               - \widetilde{\Gamma}^{ddd, \, n}_{p, n \rightarrow np, \, 110}
               - \widetilde{\Gamma}^{ddd, \, n}_{p, n \rightarrow np, \, 101}
               + \widetilde{\Gamma}^{ddd, \, n}_{p, n \rightarrow np, \, 011} +
\nonumber \\
          &&   + \widetilde{\Gamma}^{ddd, \, n}_{p, n \rightarrow np, \, 100}
               - \widetilde{\Gamma}^{ddd, \, n}_{p, n \rightarrow np, \, 010}
               - \widetilde{\Gamma}^{ddd, \, n}_{p, n \rightarrow np, \, 001}
\nonumber \\
\Gamma^{ddd}_{p, p \rightarrow np} & = &
                4 \, \widetilde{\Gamma}^{ddd, \, p}_{p, p \rightarrow np, \, 111} +
                \widetilde{\Gamma}^{ddd, \, n}_{p, p \rightarrow np, \, 111} +
                \widetilde{\Gamma}^{ddd, \, n}_{p, p \rightarrow np, \, 000} -
                \widetilde{\Gamma}^{ddd, \, n}_{p, p \rightarrow np, \, 110} +
                \widetilde{\Gamma}^{ddd, \, n}_{p, p \rightarrow np, \, 101} -
\nonumber \\
          &&   -\widetilde{\Gamma}^{ddd, \, n}_{p, p \rightarrow np, \, 011} -
                \widetilde{\Gamma}^{ddd, \, n}_{p, p \rightarrow np, \, 100} +
                \widetilde{\Gamma}^{ddd, \, n}_{p, p \rightarrow np, \, 010} -
                \widetilde{\Gamma}^{ddd, \, n}_{p, p \rightarrow np, \, 001}
\end{eqnarray}
where in the right hand side of these equations, we have put an additional
super-index ($n$ or $p$) which is the isospin projection of the $p_{1}$ particle.
As we are using different Fermi momenta for protons and neutrons, the knowledge
of the isospin projection of each particle is required. This information
about $p_{1}$ together with the transition $i, i' \rightarrow j$ is enough
to this end. The particular $\Gamma^{ddd}_{i, i' \rightarrow j}$ contribution
can be further simplified by the identification of a part of the integral in
Eqs.(\ref{rpa4}) and (\ref{rpa4p}), with the Lindhard function (details can
be found in~\cite{ba07}).

Finally, the expressions for
$\Gamma^{\alpha  \beta  \delta}_{i, i' \rightarrow j}$
with $\alpha  \beta  \delta \neq ddd$,
can be found in the Appendix. For the first order contribution
(in the nuclear residual interaction, $V^{NN}$) to the
RPA, which from now on is named as $RPA \, 1$, we can write,
\begin{eqnarray}
\label{rpafin}
\Gamma^{RPA \, 1}_{i, i' \rightarrow j} = \sum_{\alpha  \beta  \delta}
(\Gamma^{\alpha  \beta  \delta, \, l}_{i, i' \rightarrow j} +
\Gamma^{\alpha  \beta  \delta, \, u}_{i, i' \rightarrow j})
\end{eqnarray}

The final step is to show the expressions for the spectra. In the
present contribution the two-body induced decay is not evaluated.
Then, it is convenient to re-write Eqs.~(\ref{nn1f}-\ref{npp1f}), within
the first order contribution to the RPA as the only $FSI$,
\begin{eqnarray}
\label{nn1fr}
N_{n} & = & 2 \bar{\Gamma}_{n} +  \bar{\Gamma}_{p} +
\sum_{i, \, i'; \, j} N_{j \, (n)} \,
\bar{\Gamma}^{RPA \, 1}_{i, i' \rightarrow j},\\
\label{np1fr}
N_{p} & = & \bar{\Gamma}_{p} +
\sum_{i, \, i'; \, j} N_{j \, (p)} \,
\bar{\Gamma}^{RPA \, 1}_{i, i' \rightarrow j},\\
\label{nnn1fr}
N_{nn} & = & \bar{\Gamma}_{n} +
\sum_{i, \, i'; \, j} N_{j \, (nn)} \,
\bar{\Gamma}^{RPA \, 1}_{i, i' \rightarrow j},\\
\label{nnp1fr}
N_{np} & = & \bar{\Gamma}_{p} +
\sum_{i, \, i'; \, j} N_{j \, (np)} \,
\bar{\Gamma}^{RPA \, 1}_{i, i' \rightarrow j},\\
\label{npp1fr}
N_{pp} & = & 0
\end{eqnarray}
where the normalization is now, $\bar{\Gamma} \equiv \Gamma/\Gamma_{1}$.
These are already explicit expressions for the $N_{N}$ and $N_{NN}$-spectra,
once some step functions limiting the range of the momentum (or angle) of the
final particles are incorporated in Eqs.~(\ref{gamdirpi0}), (\ref{rpa4}) and
(\ref{rpa4p}). Also an additional step function is required if an
energy threshold is considered. It should be noted that the $RPA \, 1$,
together with any other
contribution in which the nuclear residual interaction appears in
first order (or in any odd order), is an interference term.
The diagrams where the nuclear residual interaction
is present an even number of times, has both interference terms
and decay widths, being the decay widths the ones where the cut is in
the middle of the diagram.
In the next section we present numerical results for these expressions.

\newpage
\section{RESULTS AND DISCUSSION}
\label{results}

In this section we present the  numerical results for the
$N_{n}$, $N_{p}$, $N_{nn}$ and $N_{np}$-spectra
using the first order contribution to the RPA.
The LDA allows us to discuss the $^{12}_{\Lambda}C$ hypernucleus.
The transition potential is represented by the exchanges
of the $\pi$, $\eta$, $K$, $\rho$, $\omega$ and $K^*$-mesons, which
formulation has been taken from \cite{pa97} and the values of the
different coupling constants and cutoff parameters
appearing in the transition
potential have been taken from \cite{na77}.
For the nuclear residual interaction, we have used the Bonn potential~\cite{ma87}
in the framework of the parametrization presented in~\cite{br96},
which contains
the exchange of $\pi$, $\rho$, $\sigma$ and $\omega$ mesons,
while the $\eta$ and $\delta$-mesons are neglected.
In implementing the LDA,
the hyperon is assumed to be in the $1s_{1/2}$ orbit of  a
harmonic oscillator well with frequency
$\hbar \omega = 10.8$ MeV.
As already stated, we have employed different values for the proton and
neutron Fermi momenta, $k_{F_n}$ and $k_{F_p}$, respectively.

To start with, let us discuss the free spectra, that is, the
spectra without final state interaction.
To this end in Eqs.~(\ref{nn1fr}-\ref{npp1fr}) all $RPA \, 1$-terms
are eliminated. For the free spectra, the INC and the microscopic scheme should
give the same result. Therefore, we make a comparison of
our values with those of the INC, in order to test our values for the spectra.
From a technical point of view, both methods
evaluate the spectra differently. While in the INC, the emitted
particles are classified according to their energies (or relative
angles), in the present microscopic model the integral which
represents the decay width is partitioned in energy (or relative
angles) regions. These two procedures are equivalent. However, some
differences show up, which are explained soon in this section.
\begin{figure}[h]
\centerline{\includegraphics[scale=0.47]{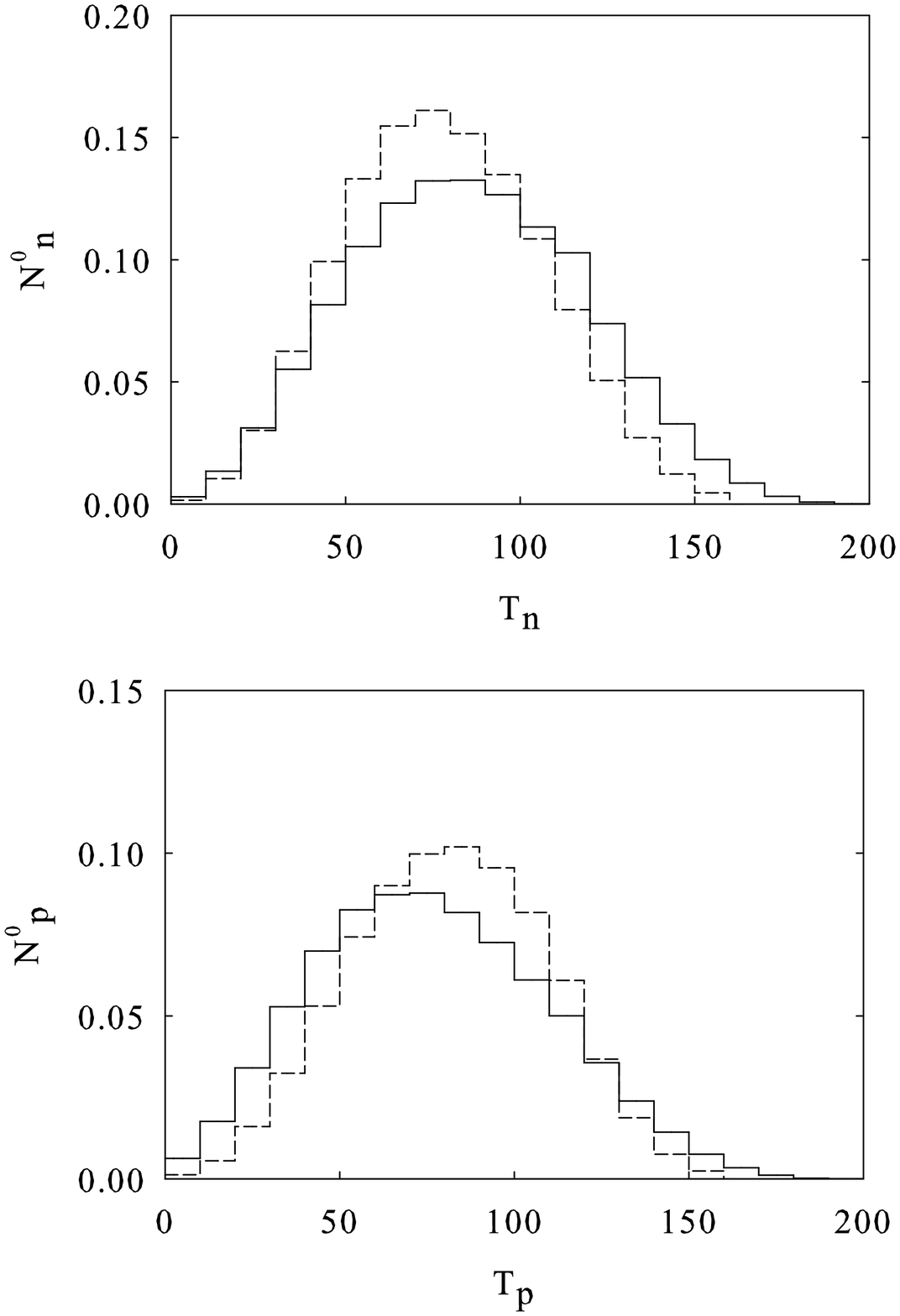}}
\caption{$N_{n}$ and $N_{p}$-free kinetic energy spectra
for $^{12}_{\Lambda}C$. The continuous
line is our result, while the dashed line represents the
INC one. The energy is given in MeV and the spectra has been
evaluated using 10 MeV steps for the kinetic energy.
We have used $\Gamma_{n}=0.267$ and $\Gamma_{p}=0.936$,
given in units of $\Gamma^0= 2.52 \cdot 10^{-6}$ eV,
where $\Gamma^0$ is the decay width of the $\Lambda$
in free space.}
\label{espnplib}
\end{figure}
In Fig.~\ref{espnplib} we present our result for
the $N_{n}$ and $N_{p}$-spectra and compare it with the corresponding
INC-result from~\cite{ba06}.
Related to the INC, we notice that our value for
the $N_{n}$-maximum is shifted towards a higher energy, while the
opposite occurs for $N_{p}$. In fact, this shift is more
marked in $N_{p}$.

\begin{figure}[h]
\centerline{\includegraphics[scale=0.47]{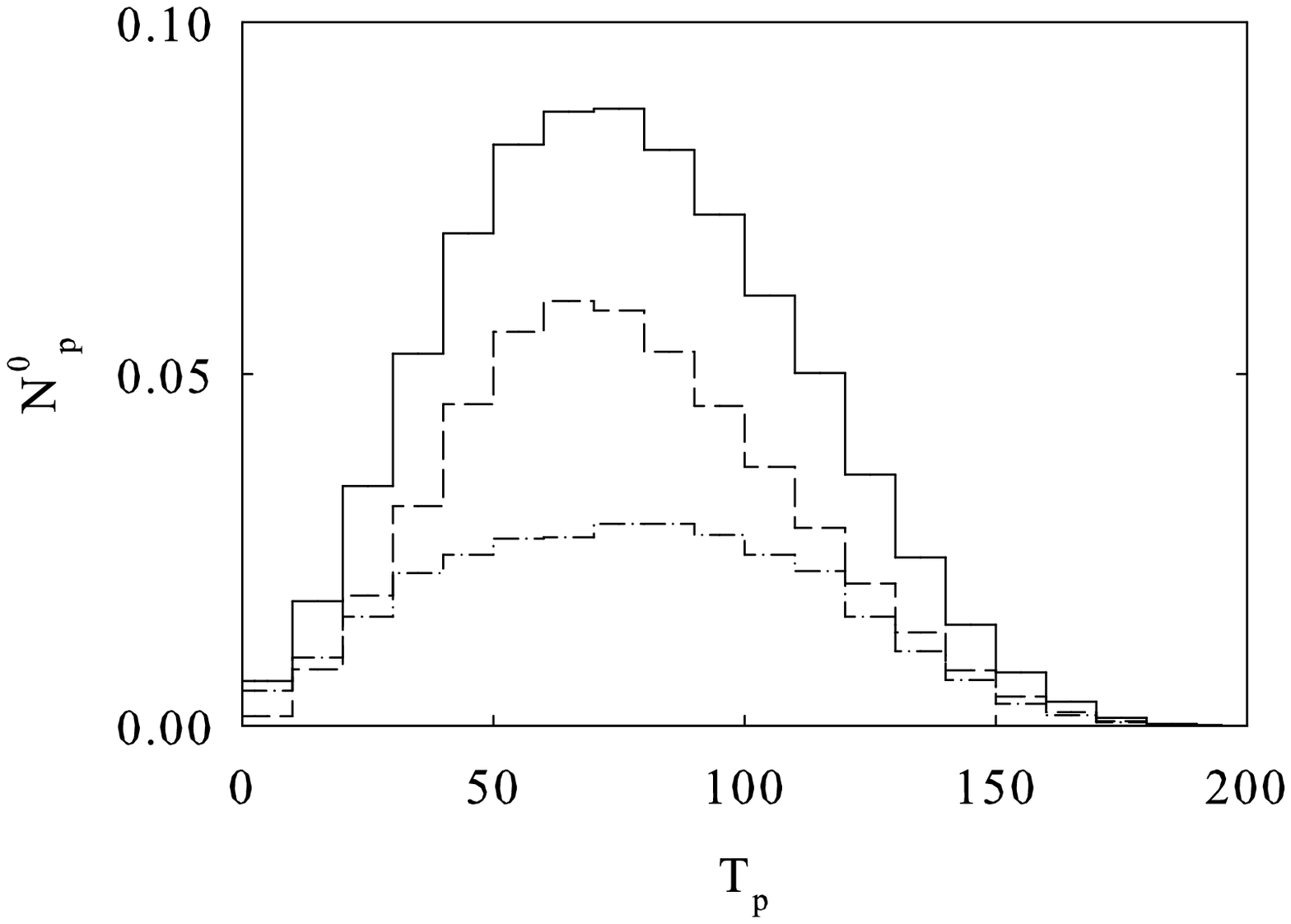}}
\caption{Isospin components to the $N_{p}$-free kinetic energy spectra.
The dashed line is the contribution where a proton is attached
to the $\Lambda$-decay vertex, the dotted-dashed line
represents the contribution where a neutron is attached
to the same vertex, while the continuous line is the total
contribution. Units are the same as in Fig.~\ref{espnplib}.}
\label{espcomp1}
\end{figure}

To understand the origin of this behavior
in Fig.~\ref{espcomp1} we have split $N_{p}$
into their two isospin components. The first isospin
component is the charge-exchange one, where
there is a proton in the $\Lambda$-vertex
(that is, in the $p_{1}$-position in Fig.~\ref{fig1rpa}).
The second is the charge-conserving contribution, which has a neutron in the just mentioned
place. The first contribution is multiplied by a factor four
(due to the isospin), while we have a
factor one for the charge-conserving term. This makes
the first contribution to be the dominant one.
Notice that the transition potential is not the same
for both contributions, because only the isovector terms
in the transition potential act for the charge-exchange contribution.
In addition, our results show that
the particle attach to the $\Lambda$-vertex is slower
than the other particle.
This fact has a simple physical interpretation. In the
expression for $\Gamma_{p}$ (or $\Gamma_{n}$), we take the principal value
for the mesons propagators in the transition potential.
As we are getting closer to the pion pole, the energy left
to the $p_{1}$-particle becomes smaller, while the
proximity to the pole, makes the contribution more important.
As mentioned above, the INC weighs this contribution with
$\Gamma_{p}$, making no distinction between the inner isospin
components. This fact is not mandatory in the INC, but the
available calculations are performed in this way.
We have implemented an isospin average to get rid
of the different isospin summation employed in the INC. We do not reproduce
these results, as they simply show the agreement between both methods.

We turn now to the inclusion of $FSI$. In the present contribution, we
have limited the $FSI$ to the first order contribution to the RPA.
In Table~I, we present our results for ${\Gamma}_{i, i' \rightarrow j}$,
within three different approximations: the first order contribution
(in the nuclear residual interaction) to the RPA ($RPA \, 1$), the same first order
term to the ring series ($ring \, 1$), and the full ring approximation taken
from~\cite{ba07}. The values quoted in this table result from
the integration of the spectra over the whole energy region.
The difference between $RPA \, 1$ and $ring \, 1$ is
the inclusion of the exchange terms. For ${\Gamma}_{n, n \rightarrow nn}$,
the exchange terms increase this quantity by $\sim 14\%$, while this percentage
is $\sim 11\%$ for ${\Gamma}_{p, p \rightarrow np}$. Although they
are different quantities, it is interesting to comment that the exchange
terms produce also an increase of $\sim 34\%$ and $\sim 30\%$ in
$\Gamma_{n}$ and $\Gamma_{p}$, respectively~\cite{ba03}. For the
two remainders terms, we have an increase (decrease) of $\sim 5\%$
($\sim 30\%$) for ${\Gamma}_{n, p \rightarrow nn}$
(${\Gamma}_{n, p \rightarrow np}$).

The next point is to make an estimation of the
accuracy of the first order RPA-contribution to account
for the full RPA series. To this end, we compare the
full ring series with $ring \, 1$. We can see that
only for ${\Gamma}_{p, p \rightarrow np}$, the
$ring \, 1$ is an adequate approximation, while
there is a strong discrepancy for all the others terms.
The reason for this behavior relays upon the inner isospin
structure of the ring (or RPA) approximation. As already stated, the
${\Gamma}_{p, p \rightarrow np}$ is dominated by
graphs where a proton is attached to the $\Lambda$-weak
decay vertex. The particle-hole bubbles (which propagates
as the ring series), are built up from a neutron particle and
a proton hole. The charge-conservation does not allow any
other particle-hole configuration. The situation is different
for ${\Gamma}_{n, n \rightarrow nn}$, ${\Gamma}_{n, p \rightarrow nn}$,
${\Gamma}_{n, p \rightarrow np}$ and
the ${\Gamma}_{p, p \rightarrow np}$-contribution with a
neutron in the $\Lambda$-vertex. As an example, we consider the
second order contribution to ${\Gamma}_{n, n \rightarrow nn}$.
This contribution has three particle-hole bubbles. The
upper and lower bubbles can be only a neutron particle-neutron hole
configuration. But the intermediate bubble can be either a
neutron particle-neutron hole or a proton particle-proton hole one.
Higher order contributions can have more complex configurations
(for details, see~\cite{ba07}). The convergence of the ring
series is fast only when the particle-hole bubbles are of the
same kind. From this analysis, it can be said that the
$RPA \, 1$ should be seen as an approximation in itself.
The ring series can be summed up to infinite
order in an easy way, but there is no
ground to neglect the exchange terms.

\begin{table}[h]
\begin{center}
\caption{Numerical results for ${\Gamma}_{i, i' \rightarrow j}$, within several
approximations.
Note that ${\Gamma}_{n, p \rightarrow j} =
{\Gamma}_{p, n \rightarrow j}$.
The $RPA \, 1$ ($ring \, 1$)-line is the first order contribution to the
$RPA$ ($ring$), while the $ring$-line represents the total ring approximation
take from~\cite{ba07}. All values are given in units of $\Gamma^{0}$.}
\vspace{1cm}
\begin{tabular}{ccccc}   \hline\hline
~~~~~~$approx.$~~~~~~ &
~~~~${\Gamma}_{n, n \rightarrow nn}$~~~~ &~2 ${\Gamma}_{n, p \rightarrow nn}$~ &
~2 ${\Gamma}_{n, p \rightarrow np}$~ &
~~~~${\Gamma}_{p, p \rightarrow np}$~~~~
\\  \hline
$RPA \, 1$   & 0.0076 & 0.0096 & 0.0064 &  0.0994  \\
$ring\, 1$ & 0.0066 & 0.0090 & 0.0068 &  0.0895   \\
$ring$   & 0.0261 & 0.0391 & 0.0371 &  0.1310  \\
\hline\hline \\
\end{tabular}
\end{center}
\end{table}

\begin{figure}[h]
\centerline{\includegraphics[scale=0.47]{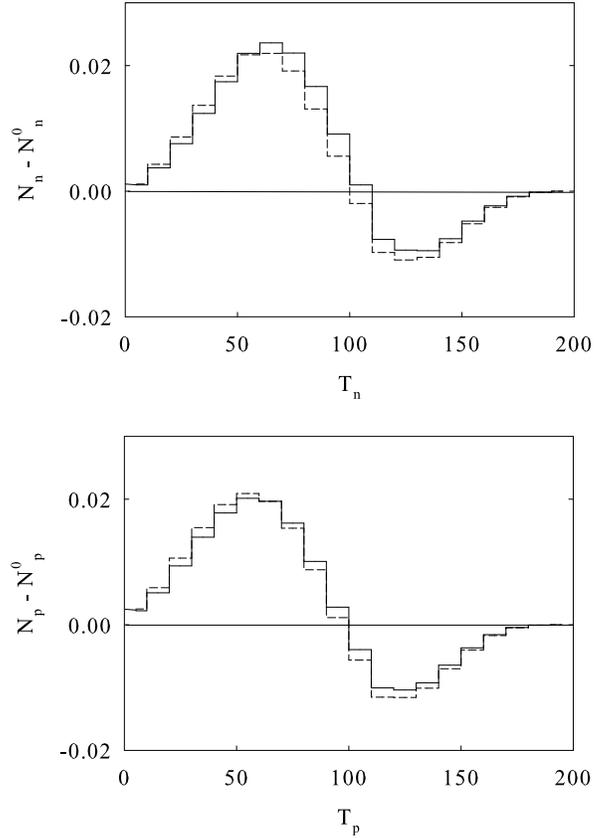}}
\caption{First order RPA contribution to the single nucleon
kinetic energy spectra
for $^{12}_{\Lambda}C$. For convenience, the free spectra has
been subtracted. The continuous line is our final $RPA \, 1$-result and
the dashed line represents the result without non-diagonal interference
terms.}
\label{espnp}
\end{figure}

\begin{figure}[h]
\centerline{\includegraphics[scale=0.47]{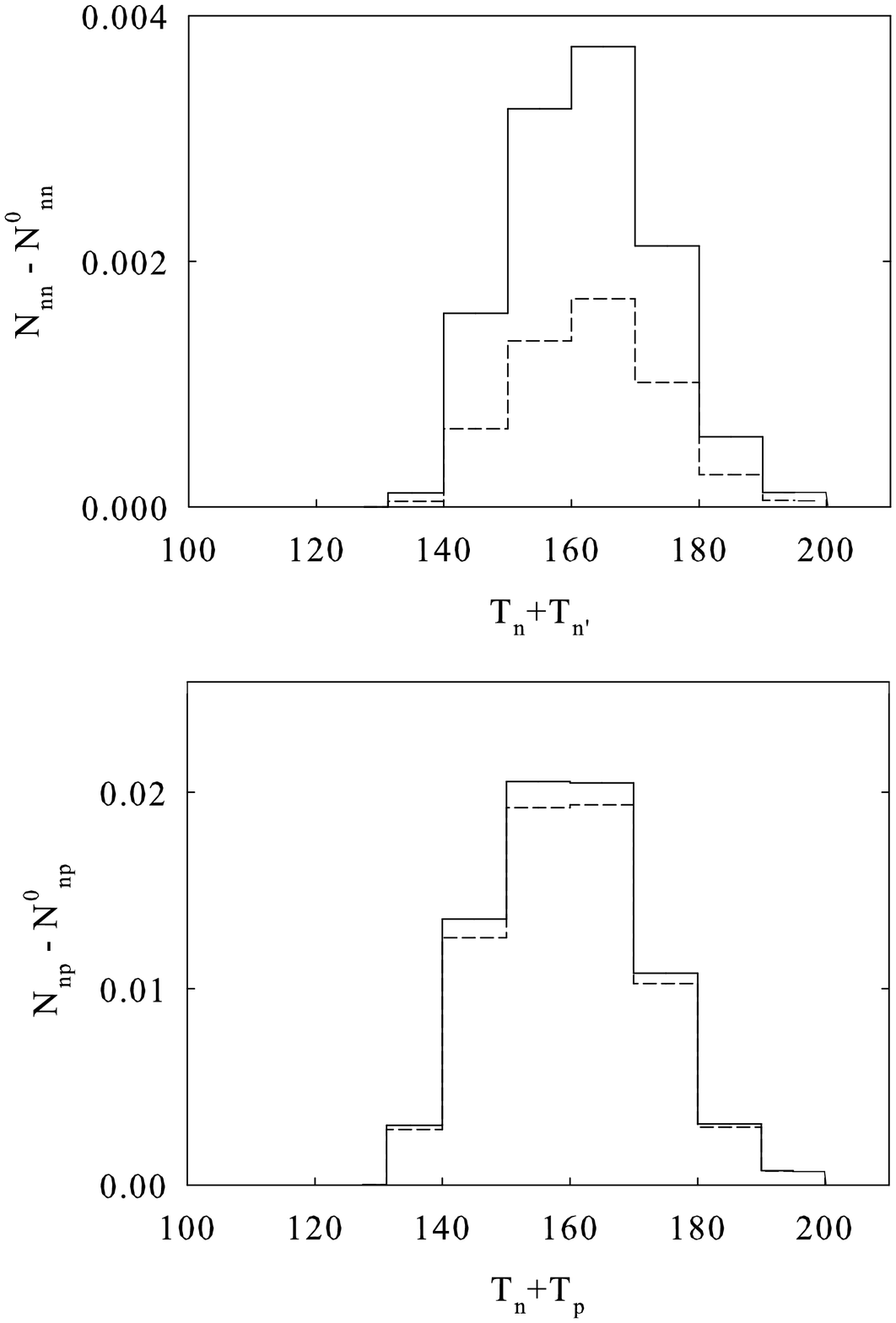}}
\caption{The same as Fig.~\ref{espnp}, but for the
double-nucleon spectra.}
\label{espnnnp}
\end{figure}

We analyze now the effect of the $FSI$ over the spectra,
where the free term has been subtracted.
This is done in Figs.~\ref{espnp} and~\ref{espnnnp},
for the $RPA \, 1$-contribution to the single and double kinetic energy
spectra, respectively. Let us recall that
within the $RPA \, 1$-model, all $FSI$ are quantum interference terms.
The ${\Gamma}_{n, n \rightarrow j}$ and ${\Gamma}_{p, p \rightarrow j}$
contributions have $i=i'$ and for convenience are called as
diagonal interference terms, while the ${\Gamma}_{n, p \rightarrow j}$-
and ${\Gamma}_{p, n \rightarrow j}$-ones
have $i \neq i'$ and are named as non-diagonal interference terms.
It should be noted that the
$\bar{\Gamma}_{i, i' \rightarrow j}$-functions can be
either positive or negative.
We present values with and without the non-diagonal
interference terms. Throughout this section we pay
much attention to study the relative importance of the non-diagonal
interference terms. Certainly, the possibility
of neglecting these terms would simplify the calculation.
In Fig.~\ref{espnp}, a typical RPA-behavior is shown .
From Table~I, together with Eqs.~(\ref{nn1fr}-\ref{npp1fr}),
the non-diagonal interference terms increase the sum of
$(N_{n}-N^{0}_{n})$ and $(N_{p}-N^{0}_{p})$, in $\sim 18\%$
and $\sim 7\%$, respectively.

In Fig.~\ref{espnnnp} a similar analysis is done for the double
kinetic energy spectra. Here, the non-diagonal interference term increases
$(N_{np}-N^{0}_{np})$ in $\sim 6\%$, while the effect over
$(N_{nn}-N^{0}_{nn})$ is very important:
it increases the result in more than a factor of two.
To understand these factors, we refer again to Table~I.
Among all the ${\Gamma}_{i, i' \rightarrow j}$,
${\Gamma}_{p, p \rightarrow np}$ is the dominant one.
Non-diagonal interference terms
are the next contribution in magnitude, but they are one order of magnitude
smaller. Only for $N_{nn}$, the ${\Gamma}_{p, p \rightarrow np}$ contribution is not present.
In this case, the non-diagonal interference term becomes very important. Once more,
the reason for this behavior relays upon the isospin factors. A charge-exchange
isospin vertex has a factor $\sqrt{2}$, in comparison with a factor one for the
charge-conserving. And within our approximation, charge-exchange contributions are only
present in ${\Gamma}_{p, p \rightarrow np}$.

\begin{figure}[h]
\centerline{\includegraphics[scale=0.47]{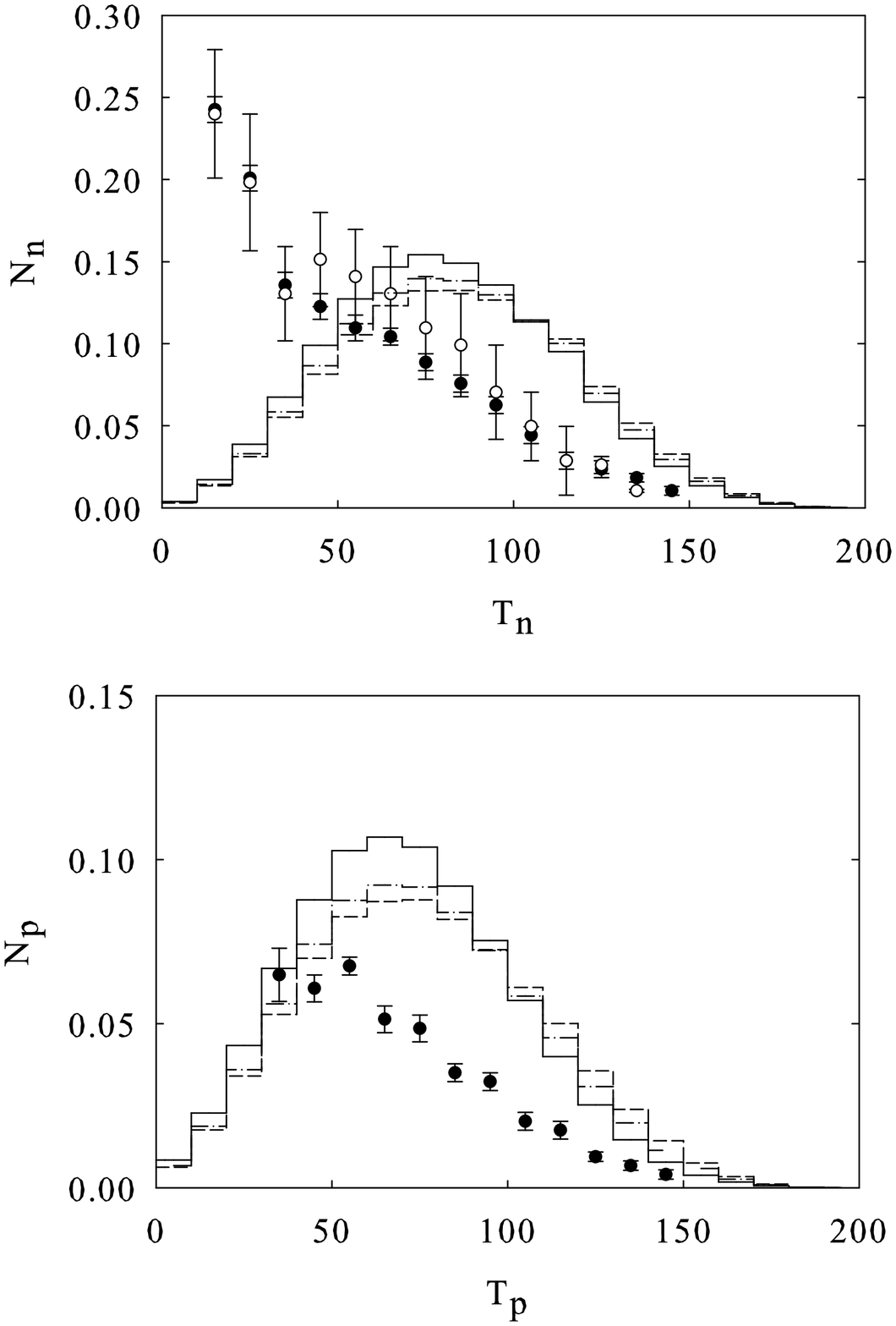}}
\caption{Final result of the $N_{n}$ and $N_{p}$
kinetic energy spectra for $^{12}_{\Lambda}C$. The continuous (dot-dashed)
line is our $RPA \, 1$-result without (with) SRC in the nuclear residual
interaction and the dashed line represents the
free result. Data are from KEK-E369~\cite{ki02} (not-filled circles)
and KEK-E508~\cite{ok04} (filled circles).}
\label{espnptot}
\end{figure}

In Fig.~\ref{espnptot}, we present our final result for the single
kinetic energy spectra $N_{n}$ and $N_{p}$ and we make a comparison
with data. In this figure, the free spectra is also shown. The
$RPA \, 1$-approximation improves the free result, as it
reduces the spectra in the high energy region, while it
produce an increase at low energies. However, it is
clear that some important $FSI$ are missing.
The comparison of our results with data in the low energy region
gives us an indication of these missing terms.
The Goldstone diagram used to evaluate the free spectra
(see Fig.~\ref{fig1rpa}), has a
two particles-one hole ($2p1h$) configuration. In fact, we have
one particle attached to the $\Lambda$-weak decay vertex, plus a
one particle-one hole bubble ($1p1h$), attached to the strong vertex. The
$RPA \, 1$ adds basically configurations restricted to $1p1h$-bubbles
(plus the corresponding exchange terms). The next $RPA$-order or
even the full $RPA$, is always restricted to this phase space.
It is likely that the required increase in phase space would be
obtained by the addition of second order
self-energy configurations.
Our knowledge from electron scattering, is that the self-energy
opens the $3p2h$-decay channel and moves intensity
towards the low energy region. The inclusion of this kind of contribution
is beyond the scope of the present contribution.
In this same figure, we present our $RPA \, 1$-results with
SRC in the nuclear residual interaction
(for details about SRC, see~\cite{ba07} and
references therein). This result is discussed soon.

\begin{figure}[h]
\centerline{\includegraphics[scale=0.47]{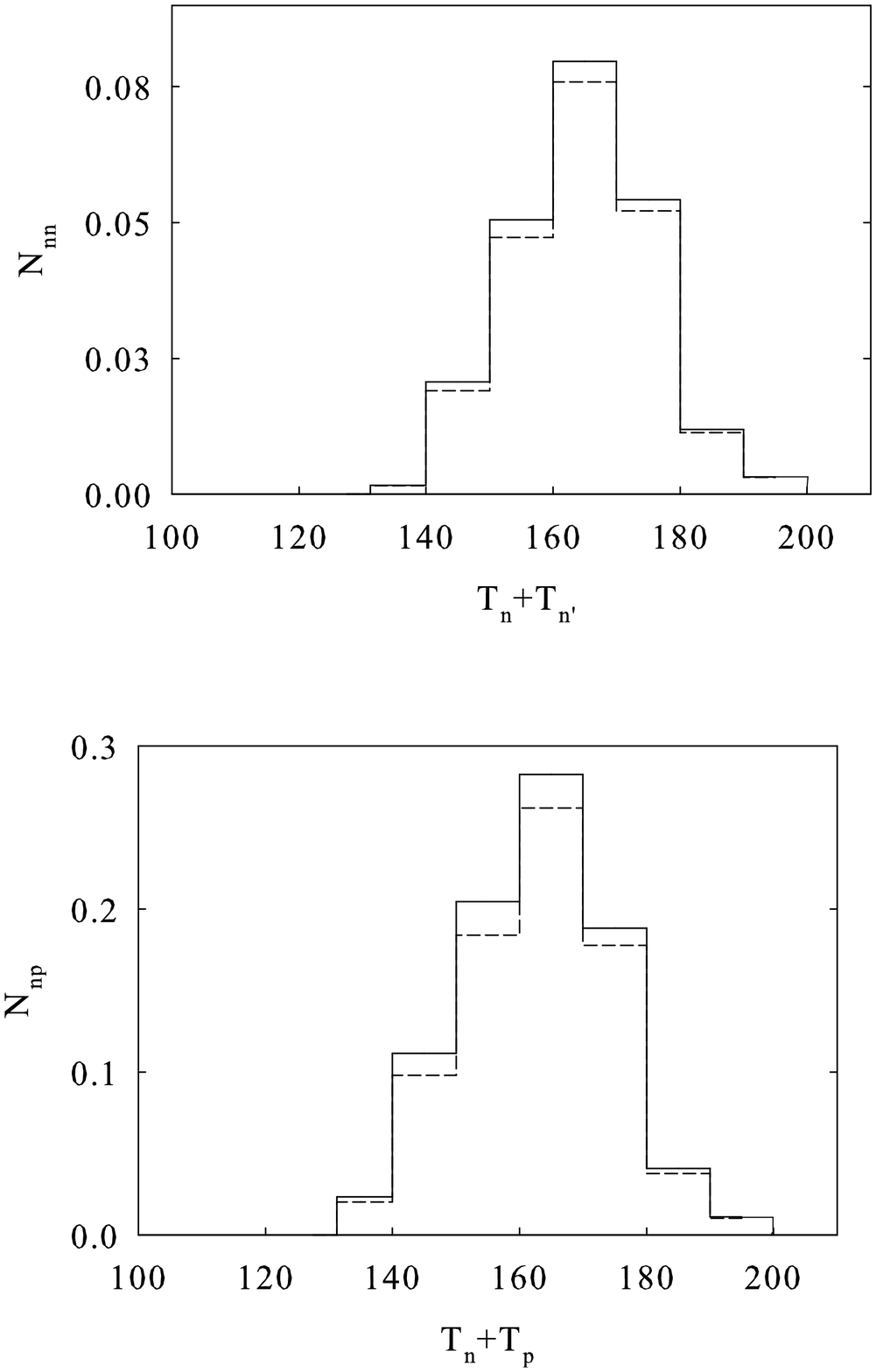}}
\caption{Final result of the $N_{nn}$ and $N_{np}$
kinetic energy spectra for $^{12}_{\Lambda}C$.
The continuous line is our $RPA \, 1$-result without SRC in the nuclear residual
interaction and the dashed line represents the
free result.}
\label{espnnnptot}
\end{figure}

\begin{figure}[h]
\centerline{\includegraphics[scale=0.47]{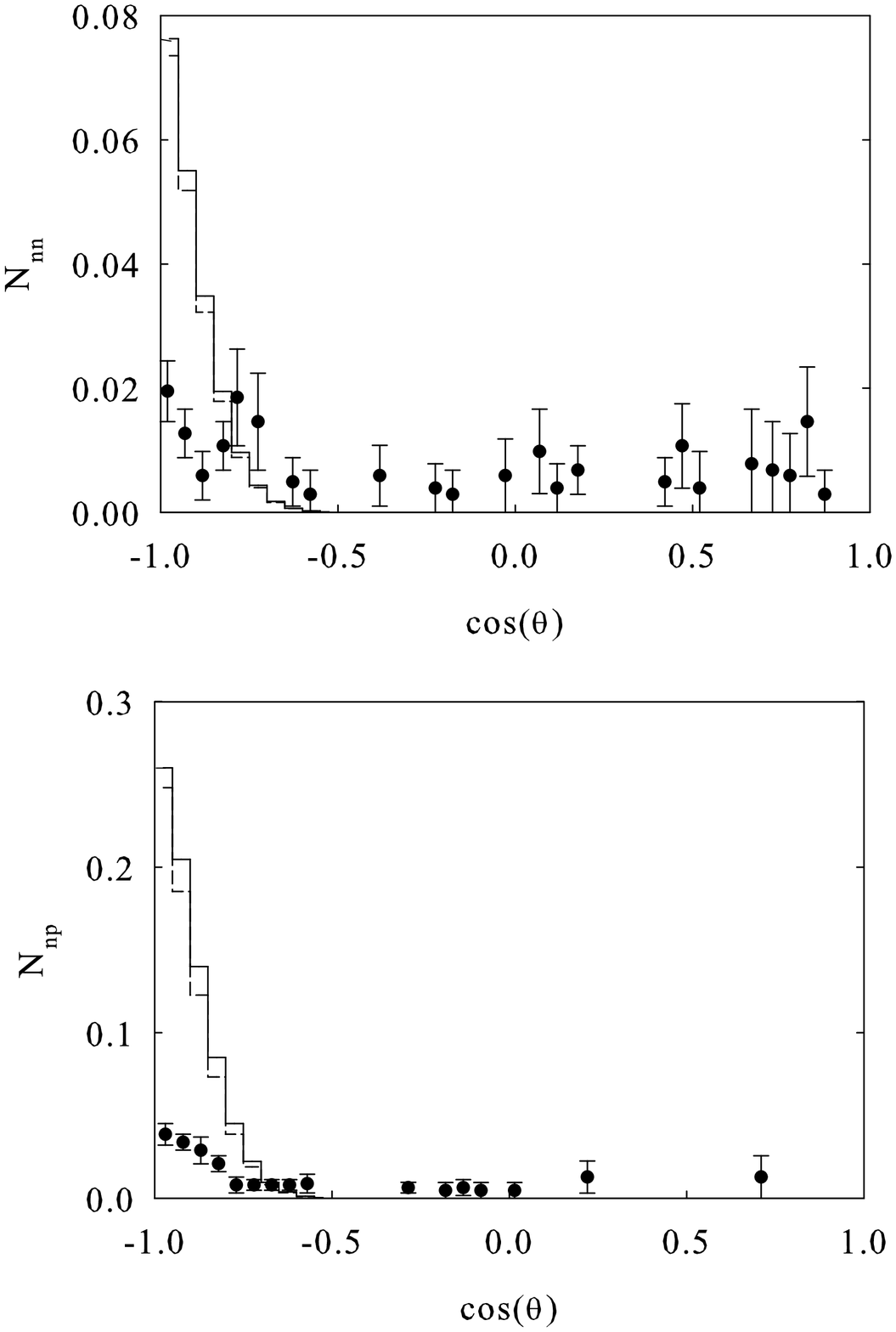}}
\caption{Opening angle distribution of $nn$ and $np$ pairs.
The continuous line is our $RPA \, 1$-result without SRC in the nuclear residual
interaction and the dashed line represents the
free result.
Data are from~\cite{kim06} }
\label{espnnnpcos}
\end{figure}

In Fig.~\ref{espnnnptot}, the double kinetic energy spectra is shown
together with the free spectra. In this case, the $RPA \, 1$-approximation
does not affect the spectra very much. Finally, in Fig.~\ref{espnnnpcos},
the opening angle distribution of $nn$ and $np$ pairs is compared with data.
Also in these figures we present both the free and the $RPA \, 1$-results.
These figures give us further evidence of the importance of including
$3p2h$-final states. The two emitted particles from the $2p1h$ final
state, emerge from the nuclei mainly with back-to-back angles.
The reason is the following: from momentum conservation,
we have $\v{k}+\v{h}_{i}=\v{p}_{1}+\v{p}_{i}$ (see Fig.~\ref{fig1rpa}).
The $\Lambda$-wave function is peaked at $|\v{k}|=0$,
while $|\v{h}_{i}|$ ranges from 0 up to $k_{F}$.
In the particular case when $|\v{k}|=|\v{h}_{i}|=0$,
we have $\v{p}_{1}=-\v{p}_{i}$, which represents
the extreme case of back-to-back kinematics.
The kinematical conditions are very different
for the $3p2h$-states, which allows the existence
of any angle between two of the three outgoing particles.

Going back to Fig.~\ref{espnnnpcos}, the data shows
intensity in forward angles, which from the theoretical point of
view would required final states like the $3p2h$-mentioned ones.
It is clear also that not only the distribution, but also the
area does not match. The origin of this discrepancy can be
understood from Fig.~\ref{espnptot}.
In accordance with the data, our spectra (in Fig.~\ref{espnnnpcos}),
has been evaluated using
an energy threshold of 30~MeV for both protons and neutrons and
our complete spectra (in Fig.~\ref{espnptot}),
has very little intensity for energies smaller than
the threshold, while data suggests that the theoretical spectra
should moves towards lower energies. In spite of this, and
due to the big error bars, we believe that
this point requires much more efforts from both
the theoretical and the experimental point of view.

In Table~II, we show our values for $N_{n}/N_{p}$ and $N_{nn}/N_{np}$,
which result from the integration of the corresponding spectra
over the whole energy region. The objective of this table is
to make a comparison of the $RPA \, 1$-result with the full ring series and with
$ring \, 1$. For this reason, no energy threshold has been
implemented, as this would affect the result, making the
analysis of the different approximations more difficult.
In the first line, we show the results without $FSI$, where
the values quoted as $dir$, refer to the first diagram in
Fig.~\ref{fig1rpa}, while the complete result is the sum
of both diagrams (direct plus exchange), in the same figure
\footnotemark{\footnotetext{To avoid confusion, it is worth to
mention that the word 'exchange' has been used for two different
contributions: the plain used of 'exchange' refers to the
Pauli exchange term, while by 'change-exchange' we indicate an
isospin component.}}.
From this table, we see that
exchange terms are important, specially for the free case.
The $RPA \, 1$ has a small influence.
In spite of this fact, it should be mentioned that the non-diagonal
interference terms represents around half of the
$FSI$-contribution. We observe that the $RPA \, 1$
decreases instead of increasing, the free value.
The opposite situation occurs with the full ring
series. This behavior is understood as a consequence of the
absence (presence) of charge-exchange terms in
${\Gamma}_{i, i' \rightarrow j}$ with $i, i' \neq p,p$,
in $RPA \, 1$ (ring approximation). Moreover, the
ring approximation leaves these ratios almost unchange
(with respect with the free result) when
the non-diagonal interference terms are not considered, while their
inclusion have more effect than in $RPA \, 1$ (or $ring \, 1$).
As a further comment from this table,
it is clear that the magnitude of the exchange terms is important
enough to object the employment of the ring series.

\begin{table}[h]
\begin{center}
\caption{The ratios $N_{n}/N_{p}$ and $N_{nn}/N_{np}$ for
$^{12}_{\Lambda}C$ hypernucleus.
Within parenthesis we present our results without
the non-diagonal interference terms. The index $dir$, indicates the
values without exchange terms. The line named as $ring$,
represents the full ring series, while the direct part
in line $RPA \, 1$, is called $ring \, 1$ as in the text. The
ring values have been taken from Tables~2 and~3 of~\cite{ba07}.}
\vspace{1cm}
\begin{tabular}{ccccc}   \hline\hline
~~~~~~~~~~~~Ref.~~~~~~~~~~~~ & ~~~~~~$N_{n}/N_{p}$~~~~~~ &
~~~~~$[N_{n}/N_{p}]^{dir}$~~~ &
~~~~~$N_{nn}/N_{np}$~~~~~& ~~~$[N_{nn}/N_{np}]^{dir}$~~
\\  \hline
LDA, no-$FSI$   & 1.571 & 1.458 &  0.285 & 0.229 \\
LDA with $RPA \, 1$  & 1.545 (1.530) &   & 0.273 (0.265) &  \\
LDA with $ring \, 1$  &  & 1.445(1.430)  &  & 0.222(0.215) \\
LDA with $ring$~\cite{ba07} &  & 1.514(1.459)  &  & 0.257(0.229) \\
\hline\hline \\
\end{tabular}
\end{center}
\end{table}

In Table~III, we show our final results for $N_{n}/N_{p}$
and $N_{nn}/N_{np}$, and compare them with data.
For $N_{n}/N_{p}$, we have
used an energy threshold of 60~MeV for both protons
and neutrons, in accordance with~\cite{ok04}; while the
results for $N_{nn}/N_{np}$ correspond to angles
between the outgoing particles, for
which $\cos(\theta)< -0.80$ and an energy threshold
of 30~MeV~\cite{bhang}. The increase in $N_{n}/N_{p}$, when
compared with the results in Table~II, is easily understood
in terms of the discussion done in Fig.~\ref{espnplib}:
The dominant charge-exchange term in $N^{0}_{p}$ makes
the maximus in the proton-induced spectra to be placed at
a smaller energy position than the neutron-induced one.
It should be mentioned that we present this table for
completeness. Our good result for $N_{n}/N_{p}$ could be
misinterpreted. The INC is used for to purposes:
the theoretical prediction of the spectra and the extraction of the so-called
experimental value for the $\Gamma_{n}/\Gamma_{p}$-ratio.
For the second point, certain coefficients are evaluated which, together with
data for $N_{n}/N_{p}$ (or $N_{nn}/N_{np}$), are used to
obtained $(\Gamma_{n}/\Gamma_{p})^{exp}$
(for details, see~\cite{ga03} and~\cite{ga04}).
Within the INC, the comparison between the
theoretical prediction for $N_{n}/N_{p}$ and $N_{nn}/N_{np}$,
with data makes much sense.
Our point of view is different: our final aim is the
reproduction of the experimental nucleon spectra. The
value for $\Gamma_{n}/\Gamma_{p}$ would be the one predicted by
our model, as the reproduction of the spectra would produce an exact match
between the theoretical and the experimental ratio, by definition.
Clearly, the determination of the ratio $\Gamma_{n}/\Gamma_{p}$ is
important because it test our knowledge of the
baryon-baryon strangeness-changing weak interactions, which is
one of the main motivations for these studies. We believe that
it is not possible to disentangle completely this issue from the
many-body problem which is intricate by itself.

\begin{table}[h]
\begin{center}
\caption{The ratios $N_{n}/N_{p}$ and $N_{nn}/N_{np}$ for
$^{12}_{\Lambda}C$ hypernucleus.
The index $diag$, indicates the values without
the quantum interference terms between the
$(\Lambda n \rightarrow nn)$ and $(\Lambda p \rightarrow np)$-amplitudes.
In accordance with data, for $N_{n}/N_{p}$ we have
used an energy threshold of 60~MeV for both protons and
neutrons, while the reported values for $N_{nn}/N_{np}$, correspond
to angles which satisfy $\cos(\theta)< -0.80$, together with
an energy threshold of 30~MeV (for both protons and
neutrons).}
\vspace{1cm}
\begin{tabular}{ccccc}   \hline\hline
~~~~~~~~~~~~Ref.~~~~~~~~~~~~ & ~~~~~~$N_{n}/N_{p}$~~~~~~ &
~~~~~$(N_{n}/N_{p})^{diag}$~~~ &
~~~~~$N_{nn}/N_{np}$~~~~~& ~~~$(N_{nn}/N_{np})^{diag}$~~
\\  \hline
LDA, no-$FSI$   & 1.791 &  &  0.279 &  \\
LDA with $RPA \, 1$  & 1.863  &  1.858 & 0.272  & 0.267 \\
KEK-E508~\cite{ok04}   & $2.00 \pm 0.09 \pm 0.14$ & &&  \\
KEK-E508~\cite{bhang} & & &  $0.40 \pm 0.09 \pm 0.04$ &\\
\hline\hline \\
\end{tabular}
\end{center}
\end{table}

As a final point for this section and
beyond the limitations in phase space in the RPA,
the nuclear residual interaction employed in the RPA, should
be discussed. To perform this calculation we have chosen
the Bonn potential where we have neglected SRC in all
results except for the ones in Fig.~\ref{espnptot}. From
this figure, we can see that
the addition of SRC reduce significantly the $RPA \, 1$-contribution.
The interaction without SRC,
has been selected to explore the RPA results under extreme conditions.
The final outcome shows us that the RPA does not alter the free
spectra very much. The addition of SRC makes this contribution
even less important. From this, we can conclude that the
$RPA \, 1$ is adequate enough to explore the RPA contribution and
from our whole analysis, it is clear that
the RPA does not represent the relevant kind of FSI required
to describe the $\Lambda$-weak decay spectra.
From the INC we know that the $FSI$ are important, which
states our problem as the finding of the most significant
diagrams to account for the $FSI$. We do not known \textit{a priori}
the answer to this question. To obtain some insight about this problem,
the RPA has been a good starting point due to it simplicity.
The search of the relevant set of diagrams has an additional
difficulty, which is the selection of the nuclear residual interaction,
together with the adjustment of some potential-parameters.
The Bonn potential is one possible choice, but the truth is
that we do not know the strong interaction in this energy-momentum region.
However, the INC provides us with a powerful tool to settle the strong interaction:
in our microscopic scheme it is easy to turn off the interference terms not
contained in the INC. Under these conditions, and once the adequate set of
diagrams are found, the INC-result can be used to fix the strong interaction.
Note that the INC does not
employ an effective strong interaction but the experimental scattering
cross section between nucleons. This would give us some confidence on
the strong interaction so adjusted.
For this particular purpose, the use of the INC is better than the data themselves,
because the theoretical model for the weak transition potential would be
the same in both the INC and in our microscopic model.
Afterwards, the interference terms should be added and a comparison with
data should be done. In our opinion, it is this whole schedule the one
which should be employed to test our knowledge of the
baryon-baryon strangeness-changing weak interactions.
The present contribution is a step forward in this direction.
In~\cite{ba07}, we have presented our scheme in general terms.
That model has been developed in the present work, but for a particular
kind of $FSI$. In a forthcoming work, the $3p2h$-decay channel which is
not contained in the RPA and can be represented by
second order self-energy contributions,
will be included. At variance with the present calculation, this
kind of $FSI$ contains both decay widths and interference
terms.

\newpage
\section{CONCLUSIONS}
\label{CONCLU}
The present work address the problem of the
theoretical interpretation of the spectra of nucleons emitted from
the non-mesonic weak decay of a $\Lambda$-hypernucleus.
To the best of our knowledge, we have presented for the first time
a scheme which deals with this issue using a microscopic formalism. We have presented
results for the single and double kinetic energy spectra and also for the
opening angle distribution of $nn$ and $np$ pairs. The $FSI$ have been
incorporated by means of the first order contribution
(in the nuclear residual interaction) to
the RPA. Our formalism naturally contains the quantum interference terms
between any pair of decay amplitudes which end in the same final state.
The quantum interference terms
appear once the $FSI$ are incorporated.
From the INC we know that the $FSI$ are important.
In a microscopic scheme there is a huge amount
of possible diagrams which constitute these $FSI$. Beforehand,
we do not known which set of diagrams is the most significant one.
The RPA has been considered as a starting point in
this analysis because it simplicity gives us some insight
on the kind of the contributions required. However,
the RPA itself does not represent an important contribution.
The comparison of our results with data suggests
that the $FSI$ should connect $2p1h$ with
$3p2h$ configurations. This kind of study is much more complex
and it is beyond the scope of the present contribution.

We have payed a particular attention to the non-diagonal
interference terms which are important within the RPA.
In fact, they are very significant in the $N_{nn}$-spectra,
while they can be neglected in $N_{np}$. The effect on
the $N_{n}$-spectra is a moderate increase and it
is also small for the $N_{p}$-spectra.
The origin of this behavior does not relays upon any
particularity of the RPA, but on some isospin factors.
We agree that only when the relevant $FSI$-diagrams are found
and incorporated, we
would be able to know if the quantum interference terms
are important or not. But from our results, it is
clear that they can not be ignored.

The existing and more successful formalism which takes care
of the same problem is the INC.
The INC is a semi-phenomenological approach, where the trajectory
of the particles emitted in the weak decay is tracked in their
way out of the nucleus. The $FSI$ are incorporated in the
INC using free path and experimental cross sections between
nucleons. Unfortunately, it is not possible to
establish a biunique relation between the processes in the
INC and in our microscopic model. The limitation of our
microscopic model is the technical difficulty in finding
the relevant $FSI$-diagrams. However, our scheme improves the
INC in two main points. The first one is the incorporation
of the quantum interference terms mentioned above. The second one refers to
the description of light hypernuclei, where the employment
of the INC is not appropriate. We have presented our results
using nuclear matter together with the LDA, which is also
a bad approximation for light hypernuclei. However, our
expressions are not limited to nuclear matter and they can
be used in a finite nucleus calculation. The incorporation
of $FSI$ in this kind of work is certainly a quite involve task.
Our aim is to describe the nuclear spectra of medium hypernuclei
where the use of nuclear matter is possible. However, once
this goal is eventually accomplished, the knowledge of
the relevant $FSI$-diagrams would encourage the evaluation of
a finite nucleus calculation for light hypernuclei.

We would like to comment now on the $\Gamma_{n}/\Gamma_{p}$-ratio.
An experimental determination of this ratio suffers from two
ambiguities: the possible contribution of two-body induced
non mesonic weak decay (which are originated from ground
state correlations, $GSC$) and the effect of $FSI$. It is worth
to mention here that these two processes should not be confused.
The two-body induced, is part of the primary decay as shown in
Eqs.~(\ref{nn10}-\ref{npp10}). After the primary decay takes place, the
strong interaction between nucleons (i.e., the $FSI$) starts acting.
The $FSI$ affects nucleons from both one- and two-body induced
decays. A final $3p2h$-state, for example, can be originated either from
a $GSC$ (without $FSI$) or from the action of $FSI$ over a one-body
induced decay. Moreover, there is an interference term between these processes,
as the initial and final state is the same. This interference term can not
be evaluated within the INC as it is a pure quantum mechanic effects.
Beyond technical difficulties, this kind of contributions are
formally part of the ${\Gamma}_{i, i' \rightarrow j}$ in
Eqs.~(\ref{nn1f}-\ref{npp1f}). This is just one example, the
theoretical problem of the inclusion of both $GSC$ and $FSI$,
is certainly a very difficult problem. Any experimental
determination of the $\Gamma_{n}/\Gamma_{p}$-ratio, assumes
a model for this theoretical problem.

Referring to the INC, it is used to predict nucleon spectra and to
extract the so-called experimental value for the $\Gamma_{n}/\Gamma_{p}$-ratio.
For the extraction of the ratio, the experimental value for
$N_{n}/N_{p}$ or $N_{nn}/N_{np}$ (together with the
corresponding error bars) is used as an input.
To put it in simple terms, a typical theoretical value for
the ratio is, $\Gamma_{n}/\Gamma_{p} \sim 0.3$, while data analyzed
by means of the INC, gives a result
$(\Gamma_{n}/\Gamma_{p})^{exp} \sim 0.4 \pm 0.1$
(we have focused on the $\cos(\theta)< -0.80$ region, for $^{12}_{\Lambda}C$).
Based on this analysis,
it is reasonable to assert that the $\Gamma_{n}/\Gamma_{p}$-puzzle
has been solved. However, one should be aware of the fact that the
experimental information used as an input in the INC is the
$N_{n}/N_{p}-$ (or $N_{nn}/N_{np}-$) ratio, where here $N_{N}$ (or
$N_{NN}$) is the integral of the spectra over energy or angle.
Clearly, this integration erase some physical information.
In fact, while the INC reproduce fairly well the
$N_{n}$-kinetic energy spectra, some discrepancies remains
for the $N_{p}$-kinetic energy spectra one.
Having in mind this element, together with the above mentioned
ambiguities,
it is our opinion that the real theoretical problem is the
reproduction of the spectra.
From the present contribution, we have indications that the
quantum interference terms should be included in the
spectra evaluation. The next step, is
to found the relevant $FSI$-diagrams. As discussed,
the $RPA \, 1$-results has given us some hints on this issue.

\section*{Acknowledgments}
I would like to thank F. Krmpoti\'{c}, G. Garbarino
and A. Ramos, for fruitful
discussions and for the critical reading of the
manuscript.
This work has been partially supported by the CONICET,
under contract PIP 6159.

\newpage

\section*{APPENDIX}
\label{APPEND}
In this Appendix we present the explicit form for
$\Gamma^{\alpha  \beta  \delta}_{i, i' \rightarrow j}$ with
$\alpha  \beta  \delta \neq ddd$. In order to simplify the expressions, all
functions which has the isospin $\tau$ ($\tau'$)-index,
carries an energy-momentum $q$ ($q'$), while the nuclear strong interaction
has an isospin index $\tau_{N}$ and has a $t$-energy momentum transfer.
Moreover, as each function is shown in a separate subsection, the
super-index $abc$ is written only when it is necessary.
\subsection{The $\Gamma^{ded}_{i, i' \rightarrow j}$-contribution}
The ${\cal S}^{ded}_{\tau' \tau_{N} \tau}(q, q', t)$ function is,
\begin{eqnarray}
\label{sded} {\cal S}^{ded}_{\tau' \, \tau_{N} \, \tau}(q,q',t) &
= & 4 {\cal V}_{C, \,\tau_{N}} \; \{ S_{\tau'}S_{\tau} +
S'_{\tau'} S'_{\tau} + 2  \, S_{V, \tau'} S_{V, \tau} + P_{C,
\tau'} P_{C, \tau}
 + \nonumber \\
& + &  3 P_{\sigma, \tau'} P_{\sigma, \tau} + P_{L, \tau'}  P_{L,
\tau} + P_{L, \tau'} P_{\sigma, \tau}+P_{\sigma, \tau'} P_{L,
\tau}
\} +  \nonumber \\
& + &  4  {\cal V}_{\sigma, \, \tau_{N}}  \; \{ 3
S_{\tau'}S_{\tau} - S'_{\tau'} S'_{\tau} - 2  \, S_{V, \tau'}
S_{V, \tau} + 3P_{C, \tau'} P_{C, \tau}
 - \nonumber \\
& - &  3 P_{\sigma, \tau'} P_{\sigma, \tau} - P_{L, \tau'}  P_{L,
\tau} - 2(P_{L, \tau'} P_{\sigma, \tau}+P_{\sigma, \tau'} P_{L,
\tau})
 \} \nonumber \\
& + & 4 {\cal V}_{L, \, \tau_{N}} \; \{ S_{\tau'}S_{\tau} -
(1+2(\v{\hat{q}} \cdot \v{\hat{t}})^{2})S'_{\tau'}S'_{\tau}
-2 (\v{\hat{q}} \cdot \v{\hat{t}})^{2} S_{V, \tau'} S_{V, \tau} + \nonumber \\
& + & (-1+2(\v{\hat{q}} \cdot \v{\hat{t}})^{2}) P_{L, \tau'} P_{L,
\tau} +
P_{C, \tau'} P_{C, \tau} -P_{\sigma, \tau'} P_{\sigma, \tau} +  \nonumber \\
&+& (-1+2(\v{\hat{q}} \cdot \v{\hat{t}})^{2}) P_{\sigma, \tau'}
P_{L, \tau}
 \}
\end{eqnarray}
where $q=k-p_{1}=q'$ and $t=h_{i}-h_{i'}$ .
\begin{eqnarray}
\label{decded}
\Gamma^{l}_{n, n \rightarrow nn} & = &
                \widetilde{\Gamma}^{l, \, n}_{n, n \rightarrow nn, \, 111} +
                \widetilde{\Gamma}^{l, \, n}_{n, n \rightarrow nn, \, 000} +
                \widetilde{\Gamma}^{l, \, n}_{n, n \rightarrow nn, \, 110} +
                \widetilde{\Gamma}^{l, \, n}_{n, n \rightarrow nn, \, 101} +
                \widetilde{\Gamma}^{l, \, n}_{n, n \rightarrow nn, \, 011} +
\nonumber \\
          &&
              + \widetilde{\Gamma}^{l, \, n}_{n, n \rightarrow nn, \, 100} +
                \widetilde{\Gamma}^{l, \, n}_{n, n \rightarrow nn, \, 010} +
                \widetilde{\Gamma}^{l, \, n}_{n, n \rightarrow nn, \, 001}
\nonumber \\
\Gamma^{u}_{n, n \rightarrow nn} & = & \Gamma^{l}_{n, n \rightarrow nn}
\nonumber \\
\Gamma^{l}_{n, p \rightarrow nn} & = & 2 (
                -\widetilde{\Gamma}^{l, \, n}_{n, p \rightarrow nn, \, 111} -
                \widetilde{\Gamma}^{l, \, n}_{n, p \rightarrow nn, \, 110} +
                \widetilde{\Gamma}^{l, \, n}_{n, p \rightarrow nn, \, 011} +
                \widetilde{\Gamma}^{l, \, n}_{n, p \rightarrow nn, \, 010})
\nonumber \\
\Gamma^{u}_{n, p \rightarrow nn} & = & 0
\nonumber \\
\Gamma^{l}_{p, n \rightarrow nn} & = & 0
\nonumber \\
\Gamma^{u}_{p, n \rightarrow nn} & = & 2 (
                -\widetilde{\Gamma}^{u, \, n}_{n, p \rightarrow nn, \, 111} +
                \widetilde{\Gamma}^{u, \, n}_{n, p \rightarrow nn, \, 110} -
                \widetilde{\Gamma}^{u, \, n}_{n, p \rightarrow nn, \, 011} +
                \widetilde{\Gamma}^{u, \, n}_{n, p \rightarrow nn, \, 010})
\nonumber \\
\Gamma^{l}_{n, p \rightarrow np} & = & 0
\nonumber \\
\Gamma^{u}_{n, p \rightarrow np} & = & 2 (
                -\widetilde{\Gamma}^{u, \, n}_{n, p \rightarrow np, \, 111} -
                \widetilde{\Gamma}^{u, \, n}_{n, p \rightarrow np, \, 110} +
                \widetilde{\Gamma}^{u, \, n}_{n, p \rightarrow np, \, 011} +
                \widetilde{\Gamma}^{u, \, n}_{n, p \rightarrow np, \, 010})
\nonumber \\
\Gamma^{l}_{p, n \rightarrow np} & = & 2 (
                -\widetilde{\Gamma}^{l, \, n}_{n, p \rightarrow np, \, 111} +
                \widetilde{\Gamma}^{l, \, n}_{n, p \rightarrow np, \, 110} -
                \widetilde{\Gamma}^{l, \, n}_{n, p \rightarrow np, \, 011} +
                \widetilde{\Gamma}^{l, \, n}_{n, p \rightarrow np, \, 010})
\nonumber \\
\Gamma^{u}_{p, n \rightarrow np} & = & 0
\nonumber \\
\Gamma^{l}_{p, p \rightarrow np} & = &
                \widetilde{\Gamma}^{l, \, n}_{p, p \rightarrow np, \, 111} +
                \widetilde{\Gamma}^{l, \, n}_{p, p \rightarrow np, \, 000} -
                \widetilde{\Gamma}^{l, \, n}_{p, p \rightarrow np, \, 110} +
                \widetilde{\Gamma}^{l, \, n}_{p, p \rightarrow np, \, 101} -
\nonumber \\
          &&   -\widetilde{\Gamma}^{l, \, n}_{p, p \rightarrow np, \, 011} -
                \widetilde{\Gamma}^{l, \, n}_{p, p \rightarrow np, \, 100} +
                \widetilde{\Gamma}^{l, \, n}_{p, p \rightarrow np, \, 010} -
                \widetilde{\Gamma}^{l, \, n}_{p, p \rightarrow np, \, 001} +
\nonumber \\
        && + 4 (-\widetilde{\Gamma}^{l, \, p}_{p, p \rightarrow np, \, 111} +
                \widetilde{\Gamma}^{l, \, p}_{p, p \rightarrow np, \, 101} )
\nonumber \\
\Gamma^{l}_{p, p \rightarrow np} & = & \Gamma^{u}_{p, p \rightarrow np}
\end{eqnarray}

\subsection{The $\Gamma^{dde}_{i, i' \rightarrow j}$ and $\Gamma^{edd}_{i, i' \rightarrow j}$-contributions}
The ${\cal S}^{dde}_{\tau' \tau_{N} \tau}(q, q', t)$ function is,
\begin{eqnarray}
\label{sdde} {\cal S}^{dde}_{\tau' \, \tau_{N} \, \tau}(q,q',t) &
= & 4 {\cal V}_{C, \,\tau_{N}} \; \{ (\v{\hat{q}} \cdot
\v{\hat{q}}')(S_{\tau'} S_{\tau} + S_{\tau'} S'_{\tau} - 2  \,
S_{\tau'} S_{V, \tau}) + P_{C, \tau'} P_{C, \tau}
 + \nonumber \\
& + &  3 P_{C, \tau'} P_{\sigma, \tau} +
P_{C, \tau'}  P_{L, \tau} \} +  \nonumber \\
& + &  4  {\cal V}_{\sigma, \, \tau_{N}}  \; \{ (\v{\hat{q}} \cdot
\v{\hat{q}}')(S'_{\tau'} S'_{\tau} + S_{\tau'} S'_{\tau} + 2
(S'_{\tau'}S_{V, \tau} +S_{\tau}S_{V, \tau'}-S'_{\tau}S_{V,
\tau'}) )
 - \nonumber \\
& - &  3 P_{\sigma, \tau'} P_{\sigma, \tau} + (-1+2(\v{\hat{q}}
\cdot \v{\hat{q}}')^{2}) P_{L, \tau'} P_{L, \tau} + 3  P_{C, \tau}
P_{\sigma, \tau'}
 +  \nonumber \\
& + &  P_{C, \tau}  P_{L, \tau'} -
  ( P_{\sigma, \tau'}  P_{L, \tau}
+ P_{L, \tau'}  P_{\sigma, \tau}) \} \nonumber \\
& + & 4 {\cal V}_{L, \, \tau_{N}} \; \{ (\v{\hat{q}} \cdot
\v{\hat{q}}')(S'_{\tau'} S'_{\tau} + S_{\tau} S'_{\tau'} + 2  \,
S'_{\tau'} S_{V, \tau})
 -P_{\sigma, \tau'} P_{\sigma, \tau} + \nonumber \\
& + & (-1+2(\v{\hat{q}} \cdot \v{\hat{q}}')^{2}) P_{L, \tau'}
P_{L, \tau} +  P_{C, \tau}  P_{\sigma, \tau'}
+  \nonumber \\
& + &  P_{C, \tau}  P_{L, \tau'} +
 (-1+2(\v{\hat{q}} \cdot \v{\hat{q}}')^{2}) P_{\sigma, \tau'}  P_{L, \tau}
-P_{L, \tau'}  P_{\sigma, \tau}) \}
\end{eqnarray}
where $q=k-p_{1}$, $t=k-p_{1}$ and $q'=p_{1}-h_{i'}$.
\begin{eqnarray}
\label{decdde}
\Gamma^{l}_{n, n \rightarrow nn} & = &
                \widetilde{\Gamma}^{l, \, n}_{n, n \rightarrow nn, \, 111} +
                \widetilde{\Gamma}^{l, \, n}_{n, n \rightarrow nn, \, 000} +
                \widetilde{\Gamma}^{l, \, n}_{n, n \rightarrow nn, \, 110} +
                \widetilde{\Gamma}^{l, \, n}_{n, n \rightarrow nn, \, 101} +
                \widetilde{\Gamma}^{l, \, n}_{n, n \rightarrow nn, \, 011} +
\nonumber \\
          &&
              + \widetilde{\Gamma}^{l, \, n}_{n, n \rightarrow nn, \, 100} +
                \widetilde{\Gamma}^{l, \, n}_{n, n \rightarrow nn, \, 010} +
                \widetilde{\Gamma}^{l, \, n}_{n, n \rightarrow nn, \, 001}
\nonumber \\
\Gamma^{u}_{n, n \rightarrow nn} & = & \Gamma^{l}_{n, n \rightarrow nn}
\nonumber \\
\Gamma^{l}_{n, p \rightarrow nn} & = &
                \widetilde{\Gamma}^{l, \, n}_{n, p \rightarrow nn, \, 111} +
                \widetilde{\Gamma}^{l, \, n}_{n, p \rightarrow nn, \, 000} +
                \widetilde{\Gamma}^{l, \, n}_{n, p \rightarrow nn, \, 110} -
                \widetilde{\Gamma}^{l, \, n}_{n, p \rightarrow nn, \, 101} -
                \widetilde{\Gamma}^{l, \, n}_{n, p \rightarrow nn, \, 011} -
\nonumber \\
          &&
              - \widetilde{\Gamma}^{l, \, n}_{n, p \rightarrow nn, \, 100} -
                \widetilde{\Gamma}^{l, \, n}_{n, p \rightarrow nn, \, 010} +
                \widetilde{\Gamma}^{l, \, n}_{n, p \rightarrow nn, \, 001}
\nonumber \\
\Gamma^{u}_{n, p \rightarrow nn} & = & 0
\nonumber \\
\Gamma^{l}_{p, n \rightarrow nn} & = & 0
\nonumber \\
\Gamma^{u}_{p, n \rightarrow nn} & = & 2 (
                - \widetilde{\Gamma}^{u, \, n}_{p, n \rightarrow nn, \, 111} +
                 \widetilde{\Gamma}^{u, \, n}_{p, n \rightarrow nn, \, 101} -
                 \widetilde{\Gamma}^{u, \, n}_{p, n \rightarrow nn, \, 011} +
                 \widetilde{\Gamma}^{u, \, n}_{p, n \rightarrow nn, \, 001} )
\nonumber \\
\Gamma^{l}_{n, p \rightarrow np} & = & 0
\nonumber \\
\Gamma^{u}_{n, p \rightarrow np} & = &
                \widetilde{\Gamma}^{u, \, n}_{n, p \rightarrow np, \, 111} +
                \widetilde{\Gamma}^{u, \, n}_{n, p \rightarrow np, \, 000} +
                \widetilde{\Gamma}^{u, \, n}_{n, p \rightarrow np, \, 110} -
                \widetilde{\Gamma}^{u, \, n}_{n, p \rightarrow np, \, 101} -
                \widetilde{\Gamma}^{u, \, n}_{n, p \rightarrow np, \, 011} -
\nonumber \\
          &&   -\widetilde{\Gamma}^{u, \, n}_{n, p \rightarrow np, \, 100} -
                \widetilde{\Gamma}^{u, \, n}_{n, p \rightarrow np, \, 010} +
                \widetilde{\Gamma}^{u, \, n}_{n, p \rightarrow np, \, 001}
\nonumber \\
\Gamma^{l}_{p, n \rightarrow np} & = &2 (
                - \widetilde{\Gamma}^{u, \, n}_{p, n \rightarrow np, \, 111} +
                 \widetilde{\Gamma}^{u, \, n}_{p, n \rightarrow np, \, 101} -
                 \widetilde{\Gamma}^{u, \, n}_{p, n \rightarrow np, \, 011} +
                 \widetilde{\Gamma}^{u, \, n}_{p, n \rightarrow np, \, 001} )
\nonumber \\
\Gamma^{u}_{p, n \rightarrow np} & = & 0
\nonumber \\
\Gamma^{l}_{p, p \rightarrow np} & = &2 (
                - \widetilde{\Gamma}^{l, \, n}_{p, p \rightarrow np, \, 111} -
                 \widetilde{\Gamma}^{l, \, n}_{p, p \rightarrow np, \, 101} +
                 \widetilde{\Gamma}^{l, \, n}_{p, p \rightarrow np, \, 011} +
                 \widetilde{\Gamma}^{l, \, n}_{p, p \rightarrow np, \, 001} ) +
\nonumber \\
          &&+ 4 (-\widetilde{\Gamma}^{l, \, p}_{p, p \rightarrow np, \, 111} +
                \widetilde{\Gamma}^{l, \, p}_{p, p \rightarrow np, \, 110})
\nonumber \\
\Gamma^{u}_{p, p \rightarrow np} & = & \Gamma^{l}_{p, p \rightarrow np}
\end{eqnarray}

For $\Gamma^{edd}_{i, i' \rightarrow j}$ we have,
$\Gamma^{edd, \, l}_{i, i' \rightarrow j} + \Gamma^{edd, \, u}_{i, i' \rightarrow j}=
\Gamma^{dde, \, l}_{i, i' \rightarrow j} + \Gamma^{dde, \, u}_{i, i' \rightarrow j}$.

\subsection{The $\Gamma^{ede}_{i, i' \rightarrow j}$-contribution}
The ${\cal S}^{ede}_{\tau' \tau_{N} \tau}(q, q', t)$ function is,
\begin{eqnarray}
\label{sede} {\cal S}^{ede}_{\tau' \, \tau_{N} \, \tau}(q,q',t) &
= & 2 {\cal V}_{C, \,\tau_{N}} \; \{ (\v{\hat{q}} \cdot
\v{\hat{q}}')(S_{\tau'} S_{\tau} + S'_{\tau'} S'_{\tau}+
S_{\tau'}S'_{\tau}+ S_{\tau}S'_{\tau'}+
4  \, S_{V, \tau'} S_{V, \tau}- \nonumber \\
&-& 2 (S_{\tau'} S_{V, \tau}+ S_{\tau} S_{V, \tau'}+S'_{\tau'}
S_{V, \tau}+ S'_{\tau} S_{V, \tau'}))+P_{C, \tau'} P_{C, \tau} +9
P_{\sigma, \tau'} P_{\sigma, \tau}
 + \nonumber \\
& + &  P_{L, \tau'} P_{L, \tau} + 3(P_{\sigma, \tau'}  P_{C,
\tau}+ P_{C, \tau'}  P_{\sigma, \tau}) +
P_{C, \tau'}  P_{L, \tau}+ P_{L, \tau'}  P_{C, \tau} +\nonumber \\
& + & 3(P_{\sigma, \tau'}  P_{L, \tau}+ P_{L, \tau'}  P_{\sigma, \tau}) \} +  \nonumber \\
& + &  2 {\cal V}_{\sigma, \, \tau_{N}}  \; \{ (\v{\hat{q}} \cdot
\v{\hat{q}}') (3S_{\tau'}S_{\tau} + 3S'_{\tau'} S'_{\tau}-
(S_{\tau'}S'_{\tau}+ S_{\tau}S'_{\tau'})+
4  \, S_{V, \tau'} S_{V, \tau}+ \nonumber \\
&+& S_{\tau'} S_{V, \tau}+ S_{\tau} S_{V, \tau'}+S'_{\tau'} S_{V,
\tau}+ S'_{\tau} S_{V, \tau'} ) + 3P_{C, \tau'} P_{C, \tau} +3
P_{\sigma, \tau'} P_{\sigma, \tau}
 + \nonumber \\
& + &  (-1+4(\v{\hat{q}} \cdot \v{\hat{q}}')^{2})P_{L, \tau'}
P_{L, \tau} - 3(P_{\sigma, \tau'}  P_{C, \tau}+ P_{C, \tau'}
P_{\sigma, \tau}) -
\nonumber \\
& - & (P_{C, \tau'}  P_{L, \tau}+ P_{L, \tau'}  P_{C, \tau}) -
(P_{\sigma, \tau'}  P_{L, \tau}+ P_{L, \tau'}  P_{\sigma, \tau})
 \} \nonumber \\
& + & 2 {\cal V}_{L, \, \tau_{N}} \; \{ (\v{\hat{q}} \cdot
\v{\hat{q}}') (S_{\tau'}S_{\tau} + S'_{\tau'} S'_{\tau})+
(-\v{\hat{q}} \cdot \v{\hat{q}}'+ 2(\v{\hat{q}} \cdot \v{\hat{t}})
(\v{\hat{q}}' \cdot \v{\hat{t}}))(S_{\tau'}S'_{\tau}+
S_{\tau}S'_{\tau'})+
\nonumber \\
&+& 2(\v{\hat{q}} \cdot \v{\hat{t}}) (\v{\hat{q}}' \cdot
\v{\hat{t}}) (S_{V, \tau'} S_{V, \tau} + S_{\tau'} S_{V, \tau}+
S_{\tau} S_{V, \tau'}+S'_{\tau'} S_{V, \tau}+
S'_{\tau} S_{V, \tau'} )+\nonumber \\
& + & P_{C, \tau'} P_{C, \tau} +P_{\sigma, \tau'} P_{\sigma, \tau}
- (P_{\sigma, \tau'}  P_{C, \tau}+ P_{C, \tau'}  P_{\sigma, \tau})
\nonumber \\
& + & (1+4(\v{\hat{q}} \cdot \v{\hat{q}}')(\v{\hat{q}} \cdot \v{\hat{t}})
(\v{\hat{q}}' \cdot \v{\hat{t}}) -
2 (\v{\hat{q}} \cdot \v{\hat{t}})^{2}
- 2(\v{\hat{q}}' \cdot \v{\hat{t}})^{2})
P_{L, \tau'} P_{L, \tau} + \nonumber \\
& + & (-1+2(\v{\hat{t}} \cdot \v{\hat{q}}')^{2})
(P_{C, \tau'}  P_{L, \tau}+ P_{L, \tau'}  P_{C, \tau}) + \nonumber \\
& + & (-1+2(\v{\hat{t}} \cdot \v{\hat{q}})^{2}) (P_{\sigma, \tau'}
P_{L, \tau}+ P_{L, \tau'}  P_{\sigma, \tau}) \}
\end{eqnarray}
where $q=p_{1}-h_{i}$, $t=k-p_{1}$ and $q'=p_{1}-h_{i'}$.
\begin{eqnarray}
\label{decede}
\Gamma^{l}_{n, n \rightarrow nn} & = &
                \widetilde{\Gamma}^{l, \, n}_{n, n \rightarrow nn, \, 111} +
                \widetilde{\Gamma}^{l, \, n}_{n, n \rightarrow nn, \, 000} +
                \widetilde{\Gamma}^{l, \, n}_{n, n \rightarrow nn, \, 110} +
                \widetilde{\Gamma}^{l, \, n}_{n, n \rightarrow nn, \, 101} +
                \widetilde{\Gamma}^{l, \, n}_{n, n \rightarrow nn, \, 011} +
\nonumber \\
          &&
              + \widetilde{\Gamma}^{l, \, n}_{n, n \rightarrow nn, \, 100} +
                \widetilde{\Gamma}^{l, \, n}_{n, n \rightarrow nn, \, 010} +
                \widetilde{\Gamma}^{l, \, n}_{n, n \rightarrow nn, \, 001}
\nonumber \\
\Gamma^{u}_{n, n \rightarrow nn} & = & \Gamma^{l}_{n, n \rightarrow nn}
\nonumber \\
\Gamma^{l}_{n, p \rightarrow nn} & = & 2 (
              - \widetilde{\Gamma}^{l, \, n}_{n, p \rightarrow nn, \, 111} -
                \widetilde{\Gamma}^{l, \, n}_{n, p \rightarrow nn, \, 110} +
                \widetilde{\Gamma}^{l, \, n}_{n, p \rightarrow nn, \, 101} +
                \widetilde{\Gamma}^{l, \, n}_{n, p \rightarrow nn, \, 100} )
\nonumber \\
\Gamma^{u}_{n, p \rightarrow nn} & = & 0
\nonumber \\
\Gamma^{l}_{p, n \rightarrow nn} & = & 0
\nonumber \\
\Gamma^{u}_{p, n \rightarrow nn} & = & 2 (
              - \widetilde{\Gamma}^{u, \, n}_{p, n \rightarrow nn, \, 111} +
                \widetilde{\Gamma}^{u, \, n}_{p, n \rightarrow nn, \, 101} -
                \widetilde{\Gamma}^{u, \, n}_{p, n \rightarrow nn, \, 011} +
                \widetilde{\Gamma}^{u, \, n}_{p, n \rightarrow nn, \, 001} )
\nonumber \\
\Gamma^{l}_{n, p \rightarrow np} & = & 0
\nonumber \\
\Gamma^{u}_{n, p \rightarrow np} & = & 2 (-
                 \widetilde{\Gamma}^{u, \, n}_{n, p \rightarrow np, \, 111} -
                 \widetilde{\Gamma}^{u, \, n}_{n, p \rightarrow np, \, 110} +
                 \widetilde{\Gamma}^{u, \, n}_{n, p \rightarrow np, \, 101} +
                 \widetilde{\Gamma}^{u, \, n}_{n, p \rightarrow np, \, 100})
\nonumber \\
\Gamma^{l}_{p, n \rightarrow np} & = & 2 (-
                 \widetilde{\Gamma}^{l, \, n}_{p, n \rightarrow np, \, 111} +
                 \widetilde{\Gamma}^{l, \, n}_{p, n \rightarrow np, \, 101} -
                 \widetilde{\Gamma}^{l, \, n}_{p, n \rightarrow np, \, 011} +
                 \widetilde{\Gamma}^{l, \, n}_{p, n \rightarrow np, \, 001})
\nonumber \\
\Gamma^{u}_{p, n \rightarrow np} & = & 0
\nonumber \\
\Gamma^{l}_{p, p \rightarrow np} & = &  2 (
                \widetilde{\Gamma}^{l, \, n}_{p, p \rightarrow np, \, 111} -
                \widetilde{\Gamma}^{l, \, n}_{p, p \rightarrow np, \, 110} -
                \widetilde{\Gamma}^{l, \, n}_{p, p \rightarrow np, \, 011} +
                \widetilde{\Gamma}^{l, \, n}_{p, p \rightarrow np, \, 010}) +
\nonumber \\
          &&  + 4 (
                \widetilde{\Gamma}^{l, \, p}_{p, p \rightarrow np, \, 111} +
                \widetilde{\Gamma}^{l, \, p}_{p, p \rightarrow np, \, 101})
\nonumber \\
\Gamma^{u}_{p, p \rightarrow np} & = & \Gamma^{l}_{p, p \rightarrow np}
\end{eqnarray}

\subsection{The $\Gamma^{dee}_{i, i' \rightarrow j}$ and $\Gamma^{eed}_{i, i' \rightarrow j}$-contributions}
The ${\cal S}^{dee}_{\tau' \tau_{N} \tau}(q, q', t)$ function is,
\begin{eqnarray}
\label{sdee} {\cal S}^{dee}_{\tau' \, \tau_{N} \, \tau}(q,q',t) &
= & 2 {\cal V}_{C, \,\tau_{N}} \; \{ (\v{\hat{q}} \cdot
\v{\hat{q}}') ((S_{\tau'} + S'_{\tau})( S_{\tau} + S'_{\tau'})-
2(S_{\tau} S_{V, \tau'} + S_{\tau'} S_{V, \tau})+ \nonumber \\
& + & 2(S'_{\tau} S_{V, \tau'} + S'_{\tau'} S_{V, \tau})) + P_{C,
\tau'} P_{C, \tau} - 3 P_{\sigma, \tau'} P_{\sigma, \tau}
 + \nonumber \\
& + & (2 (\v{\hat{q}} \cdot \v{\hat{q}}')^{2}-1)  P_{L, \tau'}
P_{L, \tau} + 3 (P_{\sigma, \tau'}  P_{C, \tau} + P_{C, \tau'}
P_{\sigma, \tau}) +
\nonumber \\
& + & P_{C, \tau'}  P_{L, \tau} + P_{L, \tau'}  P_{C, \tau} -
(P_{\sigma, \tau'}  P_{L, \tau}
+ P_{L, \tau'}  P_{\sigma, \tau}) \} +  \nonumber \\
& + &  2  {\cal V}_{\sigma, \, \tau_{N}}  \; \{ (\v{\hat{q}} \cdot
\v{\hat{q}}') (3 S_{\tau'}S_{\tau}-
S'_{\tau'}S'_{\tau}+3S_{\tau'}S'_{\tau}
-S_{\tau}S'_{\tau'}+\nonumber \\
& + & 2 S_{\tau}S_{V, \tau'}-6 S_{\tau'}S_{V, \tau} -2
S'_{\tau}S_{V, \tau'}-S'_{\tau'}S_{V, \tau}) + 3 P_{C, \tau'}P_{C,
\tau}
 + \nonumber \\
& + &  3 P_{\sigma, \tau'} P_{\sigma, \tau} + (1-(\v{\hat{q}}
\cdot \v{\hat{q}}')^{2}) P_{L, \tau'} P_{L, \tau} + 9 P_{C, \tau'}
P_{\sigma, \tau}
- 3 P_{\sigma, \tau'}  P_{C, \tau} +  \nonumber \\
& + &  3 P_{C, \tau'}  P_{L, \tau} - P_{L, \tau'}  P_{C, \tau} +
  P_{\sigma, \tau'}  P_{L, \tau}
+ P_{L, \tau'}  P_{\sigma, \tau} \} \nonumber \\
& + & 2 {\cal V}_{L, \, \tau_{N}} \; \{ (\v{\hat{q}} \cdot
\v{\hat{q}}') (S_{\tau'}S_{\tau}+ S_{\tau'}S'_{\tau}-2
S_{\tau'}S_{V, \tau})+ (-\v{\hat{q}} \cdot \v{\hat{q}}'+2
(\v{\hat{q}} \cdot \v{\hat{t}})(\v{\hat{q}}' \cdot \v{\hat{t}}))
\nonumber \\
&\times& (S'_{\tau'}S'_{\tau}+S_{\tau}S'_{\tau'}-2S'_{\tau'}S_{V,
\tau}) + (\v{\hat{q}} \cdot \v{\hat{t}})(\v{\hat{q}}' \cdot
\v{\hat{t}}) (S_{\tau}S_{V, \tau'}-S'_{\tau}S_{V, \tau'}) +
 \nonumber \\
& + & P_{C, \tau'} P_{C, \tau} +
P_{\sigma, \tau'} P_{\sigma, \tau} + \nonumber \\
& + &
(1+4(\v{\hat{q}} \cdot \v{\hat{q}}')(\v{\hat{q}} \cdot \v{\hat{t}})
(\v{\hat{q}}' \cdot \v{\hat{t}}) -
2 (\v{\hat{q}} \cdot \v{\hat{t}})^{2}
- 2(\v{\hat{q}}' \cdot \v{\hat{t}})^{2})
P_{L, \tau'} P_{L, \tau} + \nonumber \\
& + &  3 P_{C, \tau'}  P_{\sigma, \tau} - P_{\sigma, \tau'}
P_{C,\tau} + P_{C, \tau'}  P_{L, \tau} - (1-2(\v{\hat{q}} \cdot
\v{\hat{t}})^{2}) P_{L, \tau'}  P_{C, \tau}
+  \nonumber \\
& + & (-1+4(\v{\hat{q}}' \cdot \v{\hat{t}})^{2}) P_{\sigma, \tau'}
P_{L, \tau} +(1- 2(\v{\hat{q}} \cdot \v{\hat{t}})^{2}) P_{L,
\tau'} P_{\sigma, \tau} \}
\end{eqnarray}
where $q=p_{1}-h_{i}$, $t=h_{i}-h_{i'}$ and $q'=k-p_{1}$.
\begin{eqnarray}
\label{decdee}
\Gamma^{l}_{n, n \rightarrow nn} & = &
                \widetilde{\Gamma}^{l, \, n}_{n, n \rightarrow nn, \, 111} +
                \widetilde{\Gamma}^{l, \, n}_{n, n \rightarrow nn, \, 000} +
                \widetilde{\Gamma}^{l, \, n}_{n, n \rightarrow nn, \, 110} +
                \widetilde{\Gamma}^{l, \, n}_{n, n \rightarrow nn, \, 101} +
                \widetilde{\Gamma}^{l, \, n}_{n, n \rightarrow nn, \, 011} +
\nonumber \\
          &&
              + \widetilde{\Gamma}^{l, \, n}_{n, n \rightarrow nn, \, 100} +
                \widetilde{\Gamma}^{l, \, n}_{n, n \rightarrow nn, \, 010} +
                \widetilde{\Gamma}^{l, \, n}_{n, n \rightarrow nn, \, 001}
\nonumber \\
\Gamma^{u}_{n, n \rightarrow nn} & = & \Gamma^{l}_{n, n \rightarrow nn}
\nonumber \\
\Gamma^{l}_{n, p \rightarrow nn} & = & 2 (
              - \widetilde{\Gamma}^{l, \, n}_{n, p \rightarrow nn, \, 111} -
                \widetilde{\Gamma}^{l, \, n}_{n, p \rightarrow nn, \, 110} +
                \widetilde{\Gamma}^{l, \, n}_{n, p \rightarrow nn, \, 011} +
                \widetilde{\Gamma}^{l, \, n}_{n, p \rightarrow nn, \, 010} )
\nonumber \\
\Gamma^{u}_{n, p \rightarrow nn} & = & 0
\nonumber \\
\Gamma^{l}_{p, n \rightarrow nn} & = & 0
\nonumber \\
\Gamma^{u}_{p, n \rightarrow nn} & = & 4 (
                \widetilde{\Gamma}^{u, \, n}_{p, n \rightarrow nn, \, 111} +
                \widetilde{\Gamma}^{u, \, n}_{p, n \rightarrow nn, \, 011} )
\nonumber \\
\Gamma^{l}_{n, p \rightarrow np} & = & 0
\nonumber \\
\Gamma^{u}_{n, p \rightarrow np} & = & 2 (
              - \widetilde{\Gamma}^{u, \, n}_{n, p \rightarrow np, \, 111} -
                \widetilde{\Gamma}^{u, \, n}_{n, p \rightarrow np, \, 110} +
                \widetilde{\Gamma}^{u, \, n}_{n, p \rightarrow np, \, 011} +
                \widetilde{\Gamma}^{u, \, n}_{n, p \rightarrow np, \, 010} )
\nonumber \\
\Gamma^{l}_{p, n \rightarrow np} & = & 4 (
                 \widetilde{\Gamma}^{l, \, n}_{p, n \rightarrow np, \, 111} +
                 \widetilde{\Gamma}^{l, \, n}_{p, n \rightarrow np, \, 011})
\nonumber \\
\Gamma^{u}_{p, n \rightarrow np} & = & 0
\nonumber \\
\Gamma^{l}_{p, p \rightarrow np} & = &  2 (-
                \widetilde{\Gamma}^{l, \, n}_{p, p \rightarrow np, \, 111} -
                \widetilde{\Gamma}^{l, \, n}_{p, p \rightarrow np, \, 101} +
                \widetilde{\Gamma}^{l, \, n}_{p, p \rightarrow np, \, 011} +
                \widetilde{\Gamma}^{l, \, n}_{p, p \rightarrow np, \, 001}) +
\nonumber \\
          &&  + 2 (
                \widetilde{\Gamma}^{l, \, p}_{p, p \rightarrow np, \, 111} -
                \widetilde{\Gamma}^{l, \, p}_{p, p \rightarrow np, \, 110} -
                \widetilde{\Gamma}^{l, \, p}_{p, p \rightarrow np, \, 101} +
                \widetilde{\Gamma}^{l, \, p}_{p, p \rightarrow np, \, 100})
\nonumber \\
\Gamma^{u}_{p, p \rightarrow np} & = & \Gamma^{l}_{p, p \rightarrow np}
\end{eqnarray}

For $\Gamma^{eed}_{i, i' \rightarrow j}$ we have,
$\Gamma^{eed, \, l}_{i, i' \rightarrow j} + \Gamma^{eed, \, u}_{i, i' \rightarrow j}=
\Gamma^{dee, \, l}_{i, i' \rightarrow j} + \Gamma^{dee, \, u}_{i, i' \rightarrow j}$.

\subsection{The $\Gamma^{eee}_{i, i' \rightarrow j}$-contribution}
The ${\cal S}^{eee}_{\tau' \tau_{N} \tau}(q, q', t)$ function is,
\begin{eqnarray}
\label{seee} {\cal S}^{eee}_{\tau' \, \tau_{N} \, \tau}(q,q',t) &
= & 4 {\cal V}_{C, \,\tau_{N}} \; \{ (\v{\hat{q}} \cdot
\v{\hat{q}}')(S_{\tau'} S_{\tau} + S'_{\tau'} S'_{\tau} + 2  \,
S_{V, \tau'} S_{V, \tau}) + P_{C, \tau'} P_{C, \tau}
 + \nonumber \\
& + &  3 P_{\sigma, \tau'} P_{\sigma, \tau} + (\v{\hat{q}} \cdot
\v{\hat{q}}')^{2} P_{L, \tau'} P_{L, \tau} + P_{\sigma, \tau'}
P_{L, \tau}
+ P_{L, \tau'}  P_{\sigma, \tau} \} +  \nonumber \\
& + &  4  {\cal V}_{\sigma, \, \tau_{N}}  \; \{ (\v{\hat{q}} \cdot
\v{\hat{q}}')(S_{\tau'} S'_{\tau} + S'_{\tau'} S_{\tau} -
(S_{\tau'} + S'_{\tau}) ( S_{V, \tau'} + S_{V, \tau}) )
 + \nonumber \\
& + &  6 P_{\sigma, \tau'} P_{\sigma, \tau} + (1-(\v{\hat{q}}
\cdot \v{\hat{q}}')^{2}) P_{L, \tau'} P_{L, \tau} + 3 ( P_{C,
\tau'} P_{\sigma, \tau}
+ P_{\sigma, \tau'}  P_{C, \tau}) +  \nonumber \\
& + &  P_{C, \tau'}  P_{L, \tau} + P_{L, \tau'}  P_{C, \tau} +
 2 ( P_{\sigma, \tau'}  P_{L, \tau}
+ P_{L, \tau'}  P_{\sigma, \tau}) \} \nonumber \\
& + & 4 {\cal V}_{L, \, \tau_{N}} \; \{ (\v{\hat{q}} \cdot
\v{\hat{t}}) (\v{\hat{q}}' \cdot \v{\hat{t}})(S_{\tau'}
S'_{\tau} + S'_{\tau'} S_{\tau}+2  \, S_{V, \tau'} S_{V, \tau} ) - \nonumber \\
& - &  ((\v{\hat{q}} \cdot \v{\hat{q}}') - (\v{\hat{q}} \cdot
\v{\hat{t}}) (\v{\hat{q}}' \cdot \v{\hat{t}})) (S_{\tau'} +
S'_{\tau}) ( S_{V, \tau'} + S_{V, \tau})
 + 2 P_{\sigma, \tau'} P_{\sigma, \tau} + \nonumber \\
& + &
(1-(\v{\hat{q}} \cdot \v{\hat{q}}')(\v{\hat{q}} \cdot \v{\hat{t}})
(\v{\hat{q}}' \cdot \v{\hat{t}}) -
(\v{\hat{q}} \cdot \v{\hat{q}}')^{2}
- (\v{\hat{q}} \cdot \v{\hat{t}})^{2}
- (\v{\hat{q}}' \cdot \v{\hat{t}})^{2})
P_{L, \tau'} P_{L, \tau} + \nonumber \\
& + &  P_{C, \tau'}  P_{\sigma, \tau} + P_{\sigma, \tau'}
P_{C,\tau} + (\v{\hat{q}}' \cdot \v{\hat{t}})^{2} P_{C, \tau'}
P_{L, \tau} + (\v{\hat{q}} \cdot \v{\hat{t}})^{2} P_{L, \tau'}
P_{C, \tau}
+  \nonumber \\
& + & (1- (\v{\hat{q}}' \cdot \v{\hat{t}})^{2}) P_{\sigma, \tau'}
P_{L, \tau} +(1- (\v{\hat{q}} \cdot \v{\hat{t}})^{2}) P_{L, \tau'}
P_{\sigma, \tau} \}
\end{eqnarray}
where $q=p_{1}-h_{i}$, $t=h_{i}-h_{i'}$ and $q'=p_{1}-h_{i'}$.
\begin{eqnarray}
\label{deceee}
\Gamma^{l}_{n, n \rightarrow nn} & = &
                \widetilde{\Gamma}^{l, \, n}_{n, n \rightarrow nn, \, 111} +
                \widetilde{\Gamma}^{l, \, n}_{n, n \rightarrow nn, \, 000} +
                \widetilde{\Gamma}^{l, \, n}_{n, n \rightarrow nn, \, 110} +
                \widetilde{\Gamma}^{l, \, n}_{n, n \rightarrow nn, \, 101} +
                \widetilde{\Gamma}^{l, \, n}_{n, n \rightarrow nn, \, 011} +
\nonumber \\
          &&
              + \widetilde{\Gamma}^{l, \, n}_{n, n \rightarrow nn, \, 100} +
                \widetilde{\Gamma}^{l, \, n}_{n, n \rightarrow nn, \, 010} +
                \widetilde{\Gamma}^{l, \, n}_{n, n \rightarrow nn, \, 001}
\nonumber \\
\Gamma^{u}_{n, n \rightarrow nn} & = & \Gamma^{l}_{n, n \rightarrow nn}
\nonumber \\
\Gamma^{l}_{n, p \rightarrow nn} & = & 4 (
                \widetilde{\Gamma}^{l, \, n}_{n, p \rightarrow nn, \, 111} +
                \widetilde{\Gamma}^{l, \, n}_{n, p \rightarrow nn, \, 110})
\nonumber \\
\Gamma^{u}_{n, p \rightarrow nn} & = & 0
\nonumber \\
\Gamma^{l}_{p, n \rightarrow nn} & = & 0
\nonumber \\
\Gamma^{u}_{p, n \rightarrow nn} & = & 4 (
                \widetilde{\Gamma}^{u, \, n}_{p, n \rightarrow nn, \, 111} +
                \widetilde{\Gamma}^{u, \, n}_{p, n \rightarrow nn, \, 011} )
\nonumber \\
\Gamma^{l}_{n, p \rightarrow np} & = & 0
\nonumber \\
\Gamma^{u}_{n, p \rightarrow np} & = & 4 (
                \widetilde{\Gamma}^{u, \, n}_{n, p  \rightarrow np, \, 111} +
                \widetilde{\Gamma}^{u, \, n}_{n, p  \rightarrow np, \, 110} )
\nonumber \\
\Gamma^{l}_{p, n \rightarrow np} & = & 4 (
                 \widetilde{\Gamma}^{l, \, n}_{p, n \rightarrow np, \, 111} +
                 \widetilde{\Gamma}^{l, \, n}_{p, n \rightarrow np, \, 011})
\nonumber \\
\Gamma^{u}_{p, n \rightarrow np} & = & 0
\nonumber \\
\Gamma^{l}_{p, p \rightarrow np} & = &  -
                \widetilde{\Gamma}^{l, \, n}_{p, p \rightarrow np, \, 111} +
                \widetilde{\Gamma}^{l, \, n}_{p, p \rightarrow np, \, 000} +
                \widetilde{\Gamma}^{l, \, n}_{p, p \rightarrow np, \, 110} +
                \widetilde{\Gamma}^{l, \, n}_{p, p \rightarrow np, \, 101} +
\nonumber \\
       &&     + \widetilde{\Gamma}^{l, \, n}_{p, p \rightarrow np, \, 011} -
                \widetilde{\Gamma}^{l, \, n}_{p, p \rightarrow np, \, 100} -
                \widetilde{\Gamma}^{l, \, n}_{p, p \rightarrow np, \, 010} -
                \widetilde{\Gamma}^{l, \, n}_{p, p \rightarrow np, \, 001} +
\nonumber \\
          &&  + 4 (
                \widetilde{\Gamma}^{l, \, p}_{p, p \rightarrow np, \, 111} +
                \widetilde{\Gamma}^{l, \, p}_{p, p \rightarrow np, \, 110})
\nonumber \\
\Gamma^{u}_{p, p \rightarrow np} & = & \Gamma^{l}_{p, p \rightarrow np}
\end{eqnarray}

\end{document}